\documentclass[twocolumn]{aa}  
\usepackage{natbib}
\usepackage{graphicx}
\usepackage[x11names]{xcolor}
\usepackage{supertabular} 
\usepackage[toc,page]{appendix}

\usepackage[varg]{txfonts}
\usepackage{amsmath}

\usepackage{dashbox,framed,color,ocg-p}
\title{Mapping young stellar populations towards Orion with \textit{Gaia} DR1}

\author{E. Zari \inst{1}\thanks{The data and some relevant ipython notebooks used in this paper are available at https://github.com/eleonorazari/OrionDR1}, A. G. A. Brown \inst{1},
  J. de Bruijne \inst{2},  C. F. Manara \inst{2,3}  \& P. T. de Zeeuw	\inst{1,3}
}
\institute{
{1} Leiden Observatory, Niels Bohrweg 2, 2333 CA Leiden, the Netherlands; \\
{2}  Scientific Support Office, Directorate of Science, European Space Research and Technology Center (ESA/ESTEC), Keplerlaan 1, 2201 AZ Noordwijk, The Netherlands; \\
{3} ESO, Karl-Schwarzschild-Str. 2, 85748 Garching bei München, Germany
}

\abstract{
In this work we use the first data release of the \textit{Gaia} mission to explore the three dimensional arrangement and the age ordering of the many stellar groups towards the  Orion OB association, aiming at a new classification and characterization of the stellar population not embedded in the Orion A and B molecular clouds.
We make use of the parallaxes and proper motions provided in the \textit{Tycho Gaia Astrometric Solution} (TGAS) sub-set of the \textit{Gaia} catalogue, and of the combination of \textit{Gaia} and 2MASS photometry.
In TGAS, we find evidence for the presence of a young population, at a parallax $\varpi \sim 2.65 \, \mathrm{mas}$, loosely distributed around some known clusters: 25 Ori, $\epsilon$ Ori and $\sigma$ Ori, and NGC 1980 ($\iota$ Ori) and the Orion Nebula Cluster (ONC). The low mass counterpart of this population is visible in the color-magnitude diagrams constructed by combining \textit{Gaia} G photometry and 2MASS. 
We study the density distribution of the young sources in the sky, using a Kernel Density Estimation (KDE). We find the same groups as in TGAS, and also some other density enhancements that might be related to the recently discovered Orion X group, the Orion dust ring, and to the $\lambda$ Ori complex. The maps also suggest that the 25 Ori group presents a northern elongation.
We estimate the ages of this population using a Bayesian isochronal fitting procedure, assuming a unique parallax value for all the sources, and we infer the presence of an age gradient going from 25 Ori (13-15 Myr) to the ONC (1-2 Myr). We confirm this age ordering by repeating the Bayesian fit using the Pan-STARRS1 data.
Intriguingly, the estimated ages towards the NGC 1980 cluster span a broad range of values. This can either be due to the presence of two populations coming from two different episodes of star formation or to a large spread along the line of sight of the same population. Some confusion might arise from the presence of unresolved binaries, which are not modelled in the fit, and usually mimic a younger population. Finally, we provisionally relate the stellar groups to the gas and dust features in Orion. Our results form the first step towards using the \textit{Gaia} data to unravel the complex star formation history of the Orion region in terms  of the different star formation episodes, their duration, and their effects on the surrounding interstellar medium.}

\keywords{Stars: distances - stars: formation - stars: pre-main sequence - stars: early-type}

\titlerunning{Orion DR1}

\begin{document}
\maketitle
\section{Introduction}
OB stars are not distributed randomly in the sky, but cluster in loose, unbound groups, which are usually referred to as OB associations \citep{Blaauw1964}.
In the solar vicinity, OB associations are located near star-forming regions \citep{Bally2008}, hence they are prime sites for large scale studies of star formation processes and of the effects of early-type stars on the interstellar medium. 

At the end of the last century, the data of the \textit{Hipparcos} satellite \citep{Perryman1997} allowed to characterize the stellar content and the kinematic properties of nearby OB associations, deeply changing our knowledge and understanding of the solar vicinity and the entire Gould's Belt \citep{deZeeuw1999}. The canonical methods used for OB association member identification rely on the fact that stars belonging to the same OB association share the same mean velocity (plus a small random velocity dispersion). The common space velocity is perceived as a motion of the members towards a convergent point in the sky \citep[for more details see e.g.][]{deBruijne1999, Hoogerwerf1999}.   
Unfortunately, the motion of the Orion OB association is directed primarily radially away from the Sun. For this reason the methods of membership determination using the \textit{Hipparcos} proper motions did not perform well in Orion. 

The Orion star forming region is the nearest ($d \sim 400 \, \mathrm{pc}$) giant molecular cloud complex and it is a site of active star formation, including high mass stars.
All stages of star formation can be found here, from deeply embedded protoclusters, to fully exposed OB associations \citep[e.g.][]{Brown1994, Bally2008, Briceno2008, Muench2008, DaRio2014,  Getman2014}. The different modes of star formation occurring here (isolated, distributed, and clustered) allow us to study  the effect of the environment on star formation processes in great detail. Moreover, the Orion region is an excellent nearby example of the effects that young, massive stars have on the surrounding interstellar medium. The Orion-Eridanus superbubble is an expanding structure, probably driven by the combined effects of ionizing UV radiation, stellar winds, and supernova explosions from the OB association  \citep{Ochsendorf2015, Schlafly2015}. 

The Orion OB association consists of several groups, with different ages, partially superimposed along our line of sight \citep{Bally2008} and extending over an area of $\sim 30^{\circ} \times 25^{\circ}$ (corresponding to roughly $200 \, \mathrm{pc}\times 170 \, \mathrm{pc}$). 
\cite{Blaauw1964} divided the Orion OB association into four subgroups. Orion OB1a is located Northwest of the Belt stars and has an age of about 8 to 12 Myr \citep{Brown1994}. Orion OB1b contains the Belt stars and has an age estimate ranging from 1.7 to 8 Myr \citep{Brown1994, Bally2008}. Orion OB1c \citep[estimated age from 2 to 6 Myr]{Bally2008} includes the Sword stars and  is located directly in front of the Orion Nebula, M43, and NGC 1977. Hence, it is very hard to separate the stellar populations of OB1c and OB1d, the latter corresponding to the Orion Nebula Cluster \citep[ONC, see e.g.][]{DaRio2014}. It is not clear whether the entire region is a single continuous star forming event, where Ori OB1c is the more evolved stellar population emerging from the cloud where group 1d still resides, or whether 1c and 1d represent two different star formation events \citep[see e.g.][]{Muench2008}. 
In subsequent studies, many more sub-groups have been identified, such as  25 Ori \citep{Briceno2007}, $\sigma$ Ori \citep{Walter2008} and $\lambda$ Ori \citep{Mathieu2008}. Though located in the direction of the Orion OB1a and OB1b subgroups, the $\sigma$ Ori and 25 Ori sub-groups have different kinematic properties with respect to the traditional association members \citep{Briceno2007, Jeffries2006}; the $\lambda$ Ori group \citep{Mathieu2008} formation could have been triggered by the expansion of the bubble created by Orion OB1a. Its age and distance from the center of OB1a are also similar to those of OB1c.
More recently, \cite{Alves2012} and \cite{Bouy2014}  reported the discovery of a young population of stars in the foreground of the ONC, which was however questioned by \citet{DaRio2016}, \cite{Fang2017} and \cite{Kounkel2017b}. Finally, \cite{Kubiak2016} identified a rich and young population surrounding $\epsilon$ Ori. 

In this study, we use the first \textit{Gaia} data release \citep{Brown2016, Prusti2016}, hereafter \textit{Gaia} DR1, to explore the three dimensional arrangement and the age ordering of the many stellar groups between the Sun and the Orion molecular clouds,
with the overall goal to construct a new classification and characterization of the young, non-embedded stellar population in the region.  
Our approach is based on the parallaxes provided for stars brighter than $G \sim 12 \, \mathrm{mag}$ in the \textit{Tycho-Gaia Astrometric Solution} \citep[TGAS][]{Michalik2015, Lindegren2016} sub-set of the \textit{Gaia} DR1 catalogue, and on the combination of \textit{Gaia} DR1 and 2MASS photometry. These data are briefly described in Section 2. 
We find evidence for the presence of a young (age < 20 Myr) population, loosely clustered around some known groups: 25 Ori, $\epsilon$ Ori and $\sigma$ Ori, and NGC 1980 and the ONC. We derive distances to these sub-groups and (relative) ages in Section 3. In Section 4 we use the Pan-STARRS1 photometric catalogue \citep{Chambers2016}  to confirm our age ranking. 
Our results, which we discuss in Section 5 and summarize in Section 6, are the first step in utilising \textit{Gaia} data to unveil the complex star formation history of Orion and give a general overview of the episodes and the duration of the star formation processes in the entire region.

\section{Data}
\begin{figure*}
\centering
\includegraphics[width= \hsize]{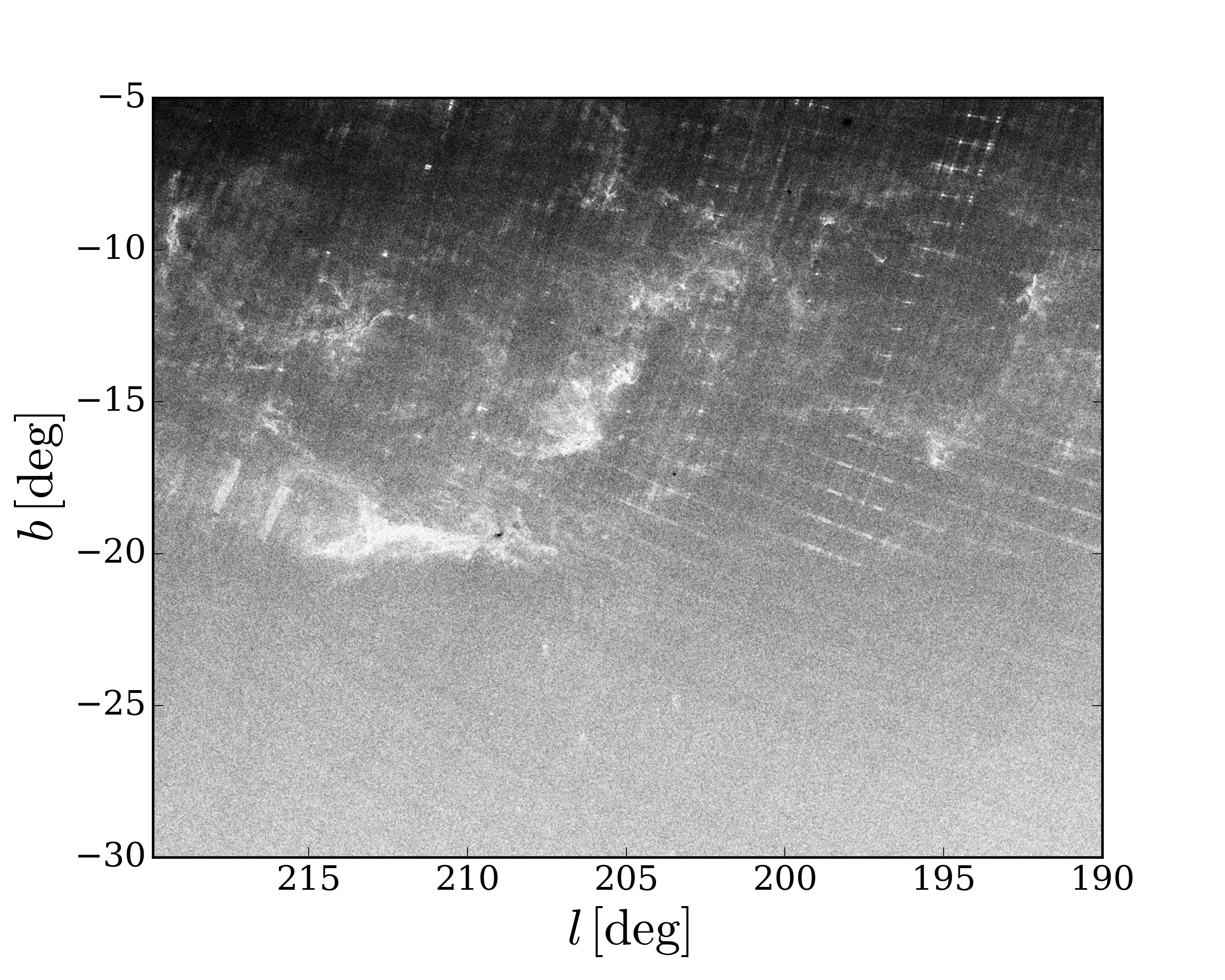}%
\caption{Sky area around the Orion constellation with the \textit{Gaia} DR1 sources selected for this study. The number of stars shown in the figure is $N = 9 926 756$. The white areas correspond to the Orion A and B molecular clouds, centred respectively at $(l, b) \sim (212, -19)$ and $(l, b) = (206, -16)$. Well visible are also the $\lambda$ Ori ring at $(l, b) \sim (196, -12)$ and Monoceros R2, at $(l, b) \sim (214, -13)$. The inclined stripes reflect the \textit{Gaia} scanning law and correspond to patches in the sky where \textit{Gaia} DR1 is highly incomplete \citep[see][]{Brown2016}.
}
\label{fig:1}
\end{figure*}

\begin{figure*}
\includegraphics[width = \hsize]{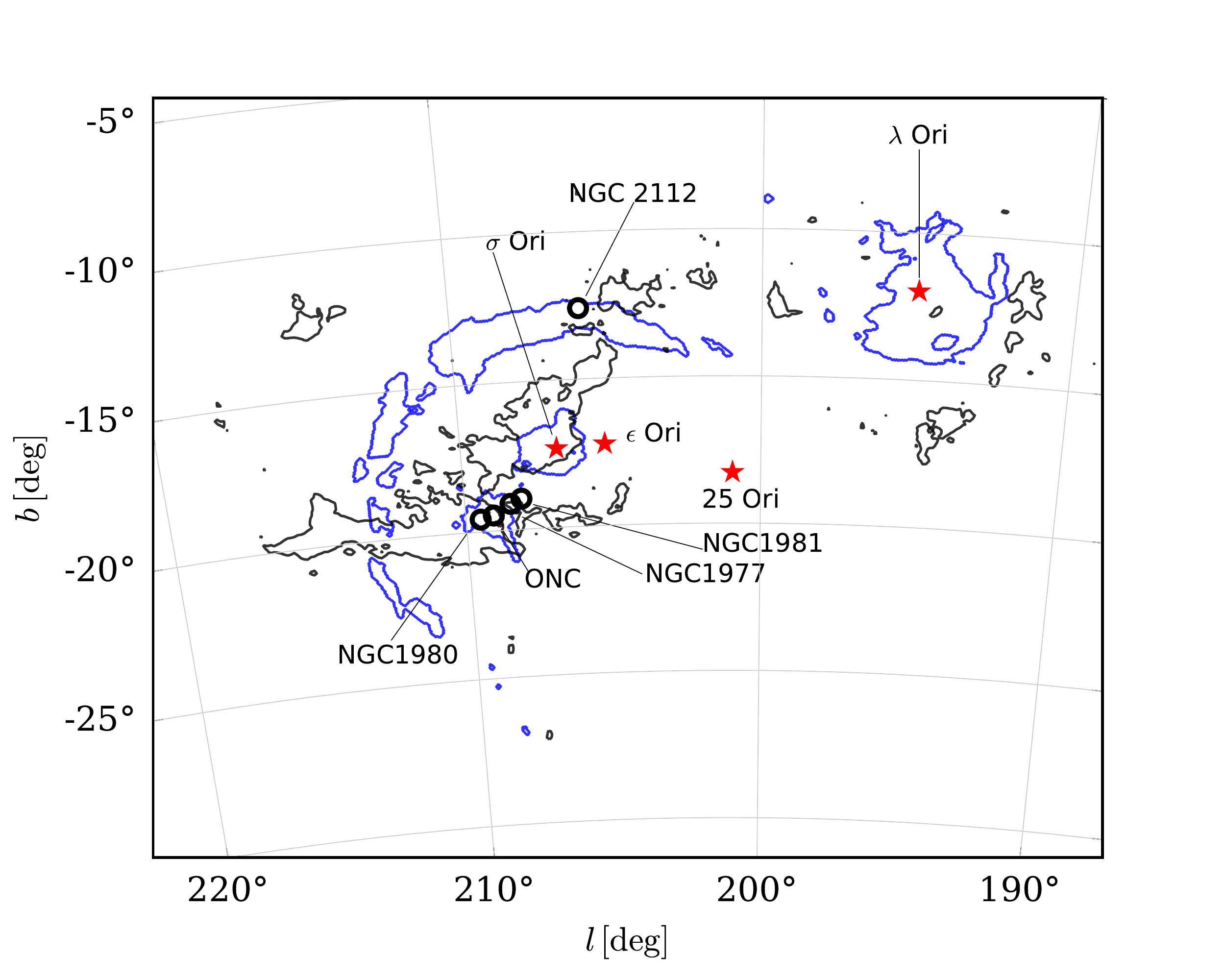}
\caption{Schematic representation of the field. The black contours correspond to the regions where $A_V > 2.5$ mag \citep{Planck2014}, while the blue contours show the $H_{\alpha}$ structures \citep{Finkbeiner2003}: Barnard's loop and the $\lambda$ Ori bubble. The positions of some known groups and stars are indicated with black circles and red stars, respectively.}
\label{fig:2}
\end{figure*}

The analysis presented in this study is based on the content of \textit{Gaia} DR1 \citep{Brown2016, vanLeeuwen2017}, complemented with the photometric data from the 2MASS catalogue \citep{Skrutskie2006} and the Pan-STARRS1 photometric catalogue \citep{Chambers2016}. 
Fig \ref{fig:1} shows the field selected for this study:
\begin{align}\label{eq:1}
190^{\circ} & <=  l <= 220^{\circ}, \nonumber \\ 
-30^{\circ}  & <= b <= -5^{\circ}.
\end{align}
We chose this field by slightly enlarging the region considered in \cite{deZeeuw1999}. 
We performed the cross-match using the \textit{Gaia} archive  (Marrese et al., in preparation). The query is reported in Appendix \ref{App:ADQL}.
In the cross-match with 2MASS, we included only the sources with photometry flag  `ph\_qual = AAA'  and we requested the angular distance of the cross-matched sources to be $< 1"$. We decided to exclude from our analysis the sources that are either young stars inside the cloud or background galaxies. We performed this filtering with a $(J-K)$ vs $(H-K_s)$ color-magnitude diagram, where extincted sources are easily identified along the reddening band. Following \cite{Alves2012},  we required that: 
\begin{align}\label{eq:cuts}
J - H & < - 1.05\,(H - K_s) + 0.97  \, \mathrm{mag}, \nonumber \\
J & < 15 \, \mathrm{mag}, \nonumber \\
H - K_s & > -0.2  \, \mathrm{mag},\, 
J - H < 0.74 \, \mathrm{mag},\,
H - K_s < 0.43 \, \mathrm{mag}.   
\end{align}
The first condition is taken as the border between non-extincted and extincted sources.
The second is meant to reject faint sources to make the selection more robust against photometric errors. The third condition excludes sources with dubious infra-red colours (either bluer or redder than main sequence stars).
The total number of \textit{Gaia} sources in the field is $N = 9,926,756$.
The number of stars resulting from the cross-match with 2MASS is $N = 5,059,068$, which further decreases to only $N = 1,450,911
$ after applying the photometric selection.
Fig. \ref{fig:2} shows a schematic representation of the field. The stellar groups relevant for this study are indicated as black empty circles and red stars. The coordinates of the stars and clusters shown are reported in Table \ref{table:1}. H$_{\alpha}$ emission \citep{Finkbeiner2003} is shown with blue contours , while dust structures \citep{Planck2014} are plotted in black. 
\begin{table}
\caption{Coordinates of the stars and clusters shown in Fig. \ref{fig:2}.}
\begin{tabular}{cc}
\hline
Name & (l, b) [deg] \\
\hline
\hline
$\lambda$ Ori  &  195, -12.0 \\
25 Ori & 201, -18.3 \\
$\epsilon$ Ori  &  205.2 -17.2 \\
$\sigma$ Ori & 206.8, -17.3 \\
NGC 1980  &  209.5, -19.6 \\
NGC 1981 &   208, -19.0 \\
NGC 1977  &  208.4, -19.1
\end{tabular}
\label{table:1}
\end{table}


\section{Orion in \textit{Gaia} DR1}\label{Sec:3}
In this section we identify and characterize the stellar population towards Orion. At first, we focus on the TGAS sub-sample and, after making a preliminary selection based on proper motions, we study the source distribution in parallax intervals. We notice the presence of an interesting concentration of sources towards the centre of the field, peaking roughly at parallax $\varpi = 2.65 \, \mathrm{mas}$ (Sec. \ref{sec:3.1}). The sources belonging to this concentration also create a sequence in the color-magnitude diagrams made combining \textit{Gaia} DR1 and 2MASS photometry (Sec. \ref{sec:3.2}). These findings prompt us to look at the entire \textit{Gaia} DR1. In the same color magnitude diagrams, we notice the presence of a young sequence, well visible between $G = 14 \, \mathrm{mag}$ and $G = 18 \, \mathrm{mag}$, which we interpret as the faint counterpart of the TGAS sequence. We make a preliminary selection of the sources belonging to the sequence, and we study their distribution in the sky, finding that they corresponded to the TGAS concentrations (Sec. \ref{sec:3.3}). We refine our selection, and finally we determine the ages of the groups we identify (Sec. \ref{sec:3.4}).

\subsection{Distances: the \textit{Tycho-Gaia} sub-sample}\label{sec:3.1}
\begin{figure*}
\includegraphics[width = \textwidth, keepaspectratio]{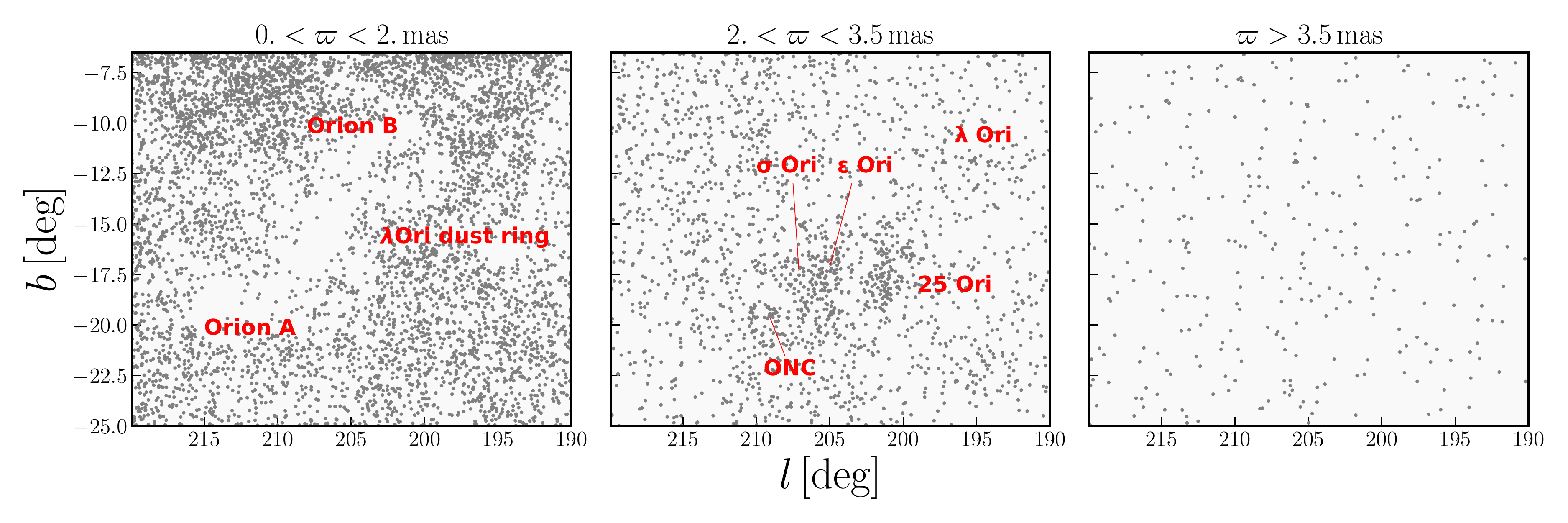}
\caption{Positions in the sky of the TGAS sources selected with Eq. \eqref{eq:3} in three different parallax intervals.
The first panel shows stars with  $0 < \varpi < 2. \, \mathrm{mas}$: the outlines of the Orion A and B molecular clouds and the $\lambda$ Ori dust ring are visible as regions with a lack of sources. The second panel shows the stars with parallax $ 2 < \varpi <3.5 \, \mathrm{mas}$. Some density enhancements are visible towards the center of the field, $(l, b) \sim (205, -18)$. The third panel shows foreground sources, with $\varpi > 3.5 \, \mathrm{mas}$.
} 
\label{fig:3}
\end{figure*}
\begin{figure}
\includegraphics[width = \hsize, keepaspectratio]{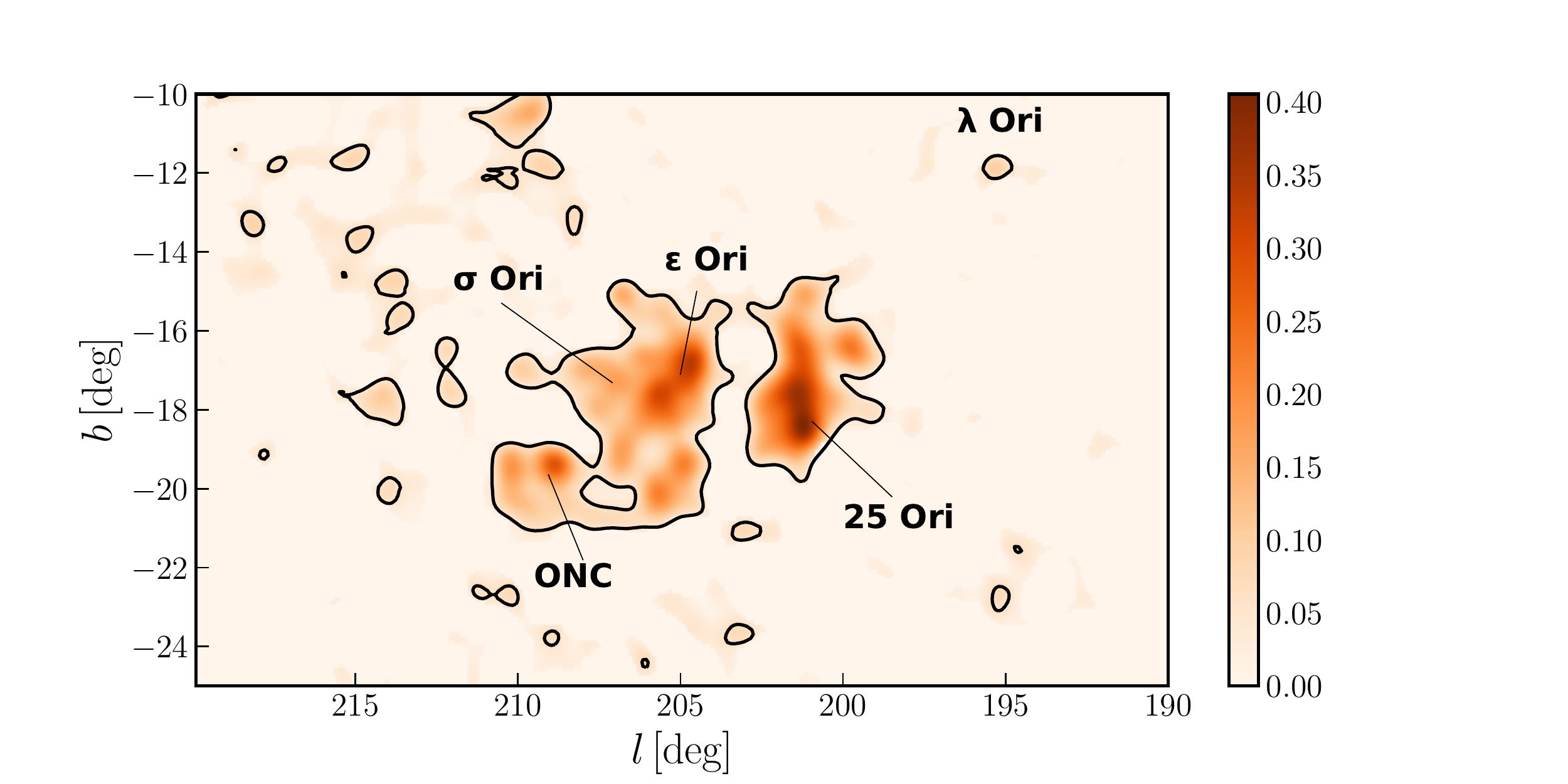}
\caption{Kernel density estimation (Gaussian Kernel with bandwidth $0.4^{\circ}$) of the TGAS sources with parallax $2 < \varpi < 3.5 \, \mathrm{mas}$.  The contours represent the $S = 3$ density levels.}
\label{fig:4}
\end{figure}

\begin{figure}
\includegraphics[width = \hsize, keepaspectratio]{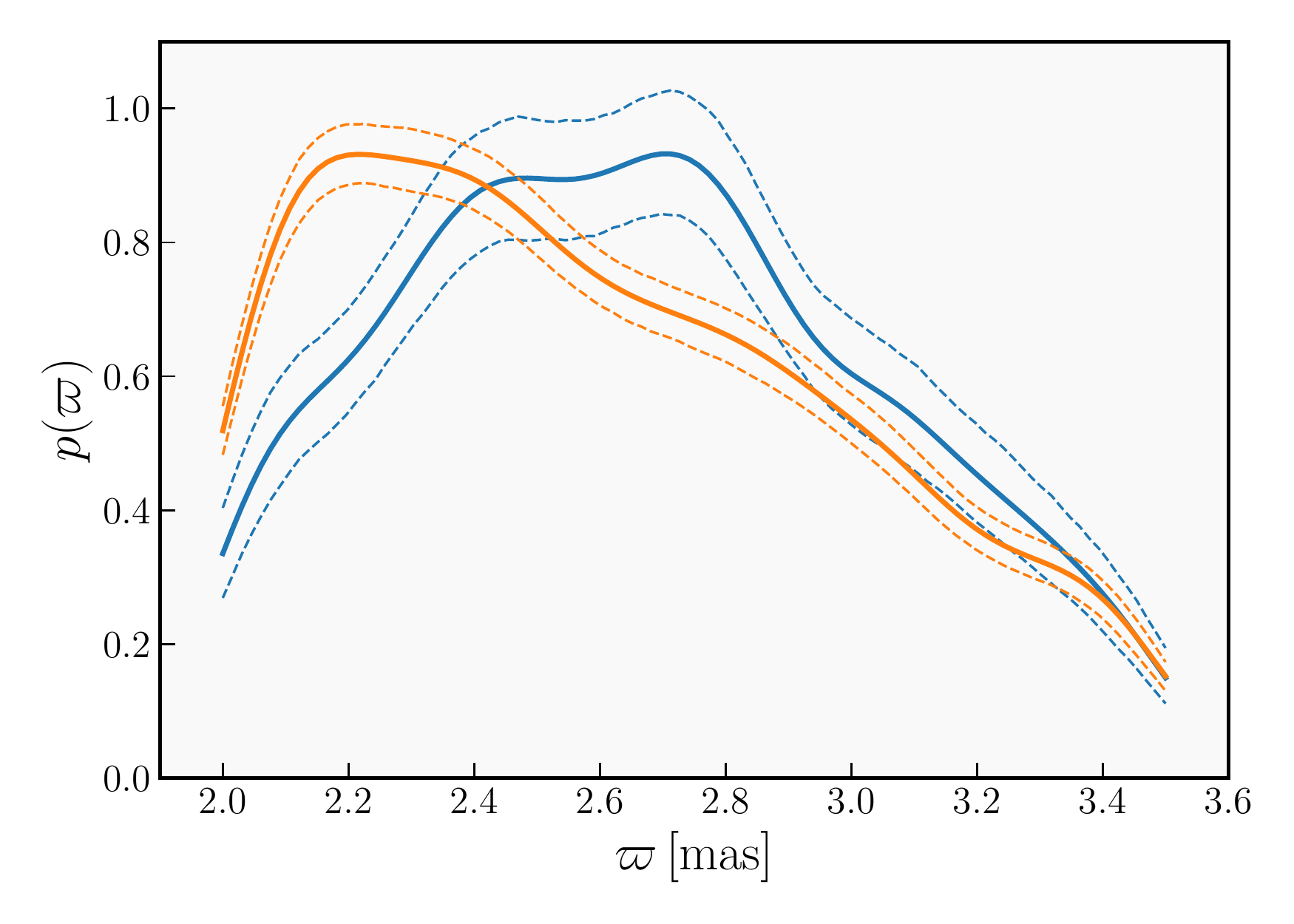}
\caption{KDE of the parallax distribution of TGAS sources with $2 < \varpi < 3.5 \, \mathrm{mas}$ (orange thick dashed line) and of the sources belonging to the density enhancements defined in the text (blue thick solid line). The fine lines represent the $5^{th}$ and $95^{th}$ percentiles, and where computed with the bootstrapping procedure described in the text. The median value of the distribution is $\varpi \sim 2.65 \, \mathrm{mas}.$}
\label{fig:5}
\end{figure}

\begin{figure*}
\includegraphics[width = \textwidth]{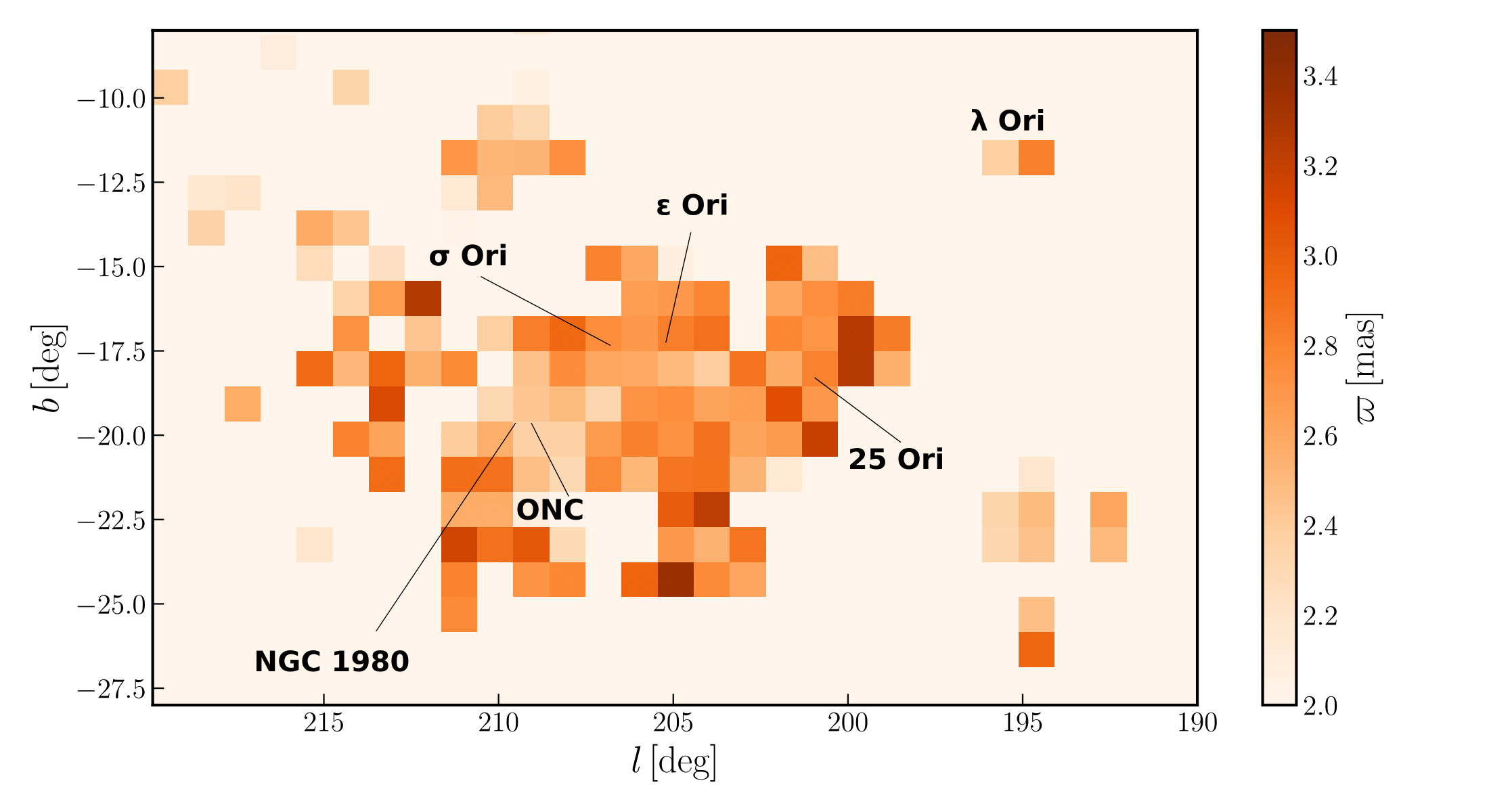}
\caption{
Median parallax of the sources within the TGAS $S = 3$ levels over bins of $1\times 1$ degrees. Along $200^{\circ} < l < 212^{\circ}$ a gradient in the parallaxes is visible, suggesting that the density enhancements visible in Fig. \ref{fig:4} have different distances, with the one associated with 25 Ori being closer than the one towards  NGC 1980. The $\lambda$ Ori group is visible at $l \sim 195^{\circ}$.
}
\label{fig:13}
\end{figure*}

Parallaxes and proper motions are available only for a sub-sample of \textit{Gaia} DR1, namely the \textit{Tycho-Gaia Astrometric Solution} \citep[TGAS][]{Michalik2015, Lindegren2016}. We consider all the TGAS sources in the field. 
Since the  motion of Orion OB1 is mostly directed radially away from the Sun, the observed proper motions are small. For this reason, a rough selection of the TGAS sources can be made requiring:
\begin{equation}\label{eq:3}
(\mu_{\alpha*} - 0.5)^2 + (\mu_{\delta}+1)^2 < 25  \,\mathrm{mas^2 \, yr^{-2}},
\end{equation}
where $\mu_{\alpha*}$ and $\mu_{\delta}$ are the proper motions in right ascension and declination. The selection above follows roughly \cite{deZeeuw1999}.
Fig. \ref{fig:3} shows the distribution  in the sky of the sources selected with Eq. \eqref{eq:3} as a function of their parallax $\varpi$, from small ($\varpi = 0 \, \mathrm{mas}$) to large parallaxes up until $\varpi = 5 \, \mathrm{mas}$ (therefore until $d = 200 \, \mathrm{pc}$).   
The outline of the Orion A and B clouds and of the $\lambda$ Ori dust ring is visible (compare with Fig. \ref{fig:1}) in the first panel, which show sources further away than $d = 500 \, \mathrm{pc}$.  This makes us confident that the sorting of sources in distance (through parallax) is correct.
The second panel in Fig. \ref{fig:3} shows stars with parallax $2 < \varpi < 3.5 \, \mathrm{mas}$, which corresponds to a distance $285 < d < 500 \, \mathrm{pc}$. 
Some source over-densities towards the center of the field, $(l, b) \sim (205^{\circ}, -18^{\circ})$, are clearly visible, and they are not due to projection effects but are indicative of real clustering in three dimensional space. 
We studied the  distribution in the sky of the sources with parallaxes $2 < \varpi < 3.5 \, \mathrm{mas}$ using a Kernel Density Estimation (KDE). The KDE is a non-parametric way to estimate the probability density function of the distribution of the sources in the sky without any assumption on their distribution. Furthermore, it smooths the contribution of each data point over a local neighbourhood and it should therefore deliver a more robust estimate of the structure of the data and its density function. We used a 
multivariate normal kernel, with isotropic  bandwidth =  $0.4^{\circ}$.  This value was chosen empirically as a good compromise between over- and under-smoothing physical density enhancements among random density fluctuations.
To avoid projection distortions, we used a metric where the distance between two points on a curved surface is determined by the haversine formula. The details of the procedure are described in Appendix C. 

To assess the significance of the density enhancements we assume that the field stars are distributed uniformly in longitude, while the source density varies in latitude. We thus average the source density over longitude along fixed latitude bins and we estimate the variance in source density using the same binning. The significance of the density enhancements is:
\begin{equation}
S(l, b) = \frac{ D(l, b) - \left\langle D(b) \right\rangle}{\sqrt{\mathrm{Var} \,(D(b))}}
\end{equation}
where $D(l, b)$ is the density estimate obtained with the KDE, $\left\langle D(b) \right\rangle$ is the average density as a function of latitude, and $\mathrm{Var}\,(D(b))$ is the variance per latitude. Fig. \ref{fig:4} shows the source probability density function, and the black contours represent the $S = 3$ levels. 
Fig. \ref{fig:5} shows the KDE of the parallax distribution
of all the sources with $2 < \varpi < 3.5 \, \mathrm{mas}$ and of those within the $S = 3$ contour levels (solid blue and orange dashed line, respectively). We used a Gaussian Kernel with bandwidth = 0.1 $\mathrm{mas}$, which is comparable to the average parallax error ($\sim 0.3 \mathrm{mas}$).
The distribution of the sources within the $S=3$ contour levels peaks at  $\varpi \sim 2.65 \, \mathrm{mas}$. This supports the notion that the stars within the density enhancements are concentrated in space.
To confirm the significance of the difference between the parallax distribution of the two samples, we performed $N = 1000$ realizations of the parallax density distribution (of both samples) by randomly sampling the single stellar parallaxes, then we computed  the $5^{th}$ and the $95^{th}$ percentiles, which are shown as fine lines in \ref{fig:5}.
Finally, we noticed that the spread in the parallax distribution ($\sim 0.5 \, \mathrm{mas}$) is larger than the typical parallax error, therefore we can hypothesize that it is due to an actual distance spread of $\sim 150 \, \mathrm{pc}$, and not only to the dispersion induced by the errors.

Fig. \ref{fig:13} shows the median parallax over bins of $1^{\circ} \times 1^{\circ}$ for the sources within the $S = 3$ levels.
The stars associated with 25 Ori have slightly larger parallaxes than those in the direction towards the ONC, which implies smaller distances from the Sun. We computed the median parallaxes in $2^{\circ}\times 2^{\circ}$ boxes centred in 25 Ori, $\epsilon$ Ori and the ONC. We obtained:
\begin{itemize}
\item 25 Ori: $\varpi = 2.81^{+0.46}_{-0.46} \,  \mathrm{mas}$  ($d \sim 355 \, \mathrm{pc}$);
\item $\epsilon$ Ori: $\varpi = 2.76^{+0.33}_{-0.35} \,  \mathrm{mas}$ ($d \sim 362 \, \mathrm{pc}$);
\item ONC: $\varpi = 2.42^{+0.2}_{-0.22} \,  \mathrm{mas}$ ($d \sim 413$),
\end{itemize}
where the quoted errors correspond to the $16^{th}$ and $84^{th}$ percentiles.

These values are consistent with the photometric distances determined by \cite{Brown1994}: $380 \pm 90 \, \mathrm{pc}$ for Ori1a; $360 \pm 70\, \mathrm{pc}$ for Ori OB1b; and $400 \pm 90 \mathrm{pc}$ for OB1c.
Using the \textit{Hipparcos} parallaxes \cite{deZeeuw1999} reported the mean distances to be:  $336 \pm 16 \, \mathrm{pc}$ for Ori OB1a; $473 \pm 33\, \mathrm{pc}$ for Ori OB1b; and $506 \pm 37 \mathrm{pc}$ for Ori OB1c. 
Distances to the Orion Nebula Cluster have been determined by, among others: \citet{Stassun2004, Hirota2007, Jeffries2007, Menten2007, Sandstrom2007, Kim2008} and \citet{Kraus2009}. These distance estimates range from $389^{+24}_{-21} \, \mathrm{pc}$ to  $437 \pm 19\, \mathrm{pc}$. The latest distance estimate was obtained by \cite{Kounkel2017a}, who found a distance of $388 \pm 5 \, \mathrm{pc}$  using radio VLBA observations of Young Stellar Objects (YSOs). Thus the TGAS distances are quite in agreement with the estimates above.

\subsection{Color magnitude diagrams}\label{sec:3.2}
\begin{figure*}
\includegraphics[width = \hsize]{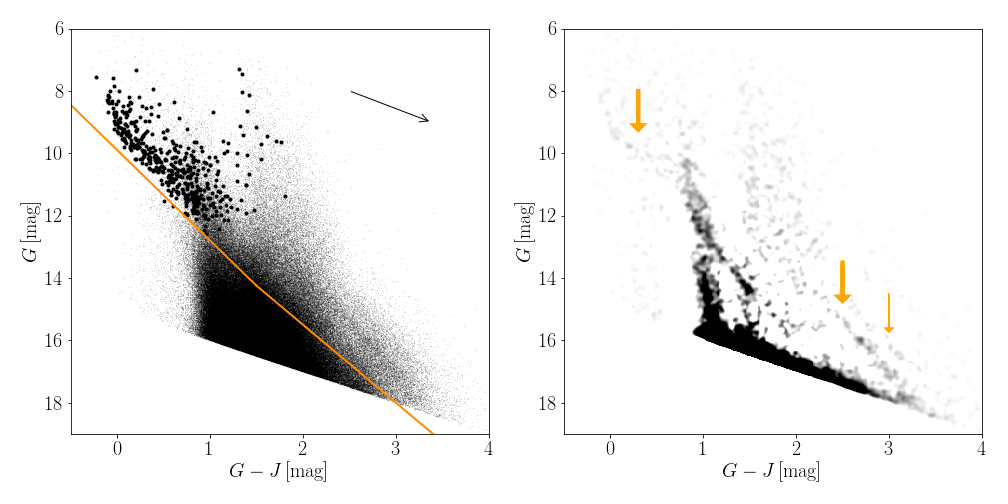}
\caption{Left: colour magnitude diagram of the \textit{Gaia} sources cross matched with 2MASS. The sources we focused on are those responsible for the dense, red sequence in the lower part of the diagram. The orange line is defined in Eq. \eqref{eq:2}, and was used to separate the bulk of the field stars from the population we intended to study. The big black points represent the sources within the TGAS $S = 3$ contour levels of Fig. \ref{fig:4}. The arrow shows the reddening vector corresponding to $A_V = 1 \, \mathrm{mag}$. Right: same color magnitude diagram as on the left, after unsharp masking. The most interesting features (bright, TGAS sequence; faint \textit{Gaia} DR1 sequence; binary sequence) are highlighted with the orange arrows.}
\label{fig:6}
\end{figure*}
We combine \textit{Gaia} and 2MASS photometry to make color-magnitude diagrams of the sources within the $S = 3$ levels defined in Fig. \ref{fig:4}. These sources define a sequence at the bright end of the color-magnitude diagram (black big dots in Fig. \ref{fig:6}, left). The spread of the sequence does not significantly change using apparent or absolute magnitudes. 
This prompts us to look further at the entire field, using the entire \textit{Gaia} DR1 catalogue to find evidence of the faint counterpart of the concentration reported in Sec. \ref{sec:3.1}. 
Fig.\ref{fig:6} (left) shows a $G$ vs. $G-J$ color magnitude diagram of the central region of the field, with coordinates:
\begin{align*}
195^{\circ} &< l < 212^{\circ}, \nonumber \\ 
-22^{\circ}  & < b < -12^{\circ}.
\end{align*}
Fig. \ref{fig:6} (right) shows the same color magnitude diagram after unsharp masking. A dense, red sequence is visible between $G = 14 \, \mathrm{mag}$ and $G = 18 \, \mathrm{mag}$. This kind of sequence \citep[also reported for example by][]{Alves2012} indicates the presence of a population of young stars. Indeed, the locus of the sequence is situated above the main sequence at the distance of Orion. 
Several basic characteristics can be inferred from the diagram: 
\begin{enumerate}
\item The density of the sequence suggests that the population is rich;
\item The sequence appears not to be significantly affected by reddening, indicating that the sources are in front of or at the edges of the clouds;
\item The dispersion of the sequence is $\sim 0.5 \,\mathrm{mag}$.
This can be due to multiple reasons, such as: the presence of unresolved binaries, the presence of groups of different ages or distances, or of field contaminants.
\end{enumerate}
Since our field is large, the number of contaminants is high. Therefore, we decided to eliminate the bulk of the field stars by requiring the following conditions to hold (orange line in Fig. \ref{fig:6} left):
\begin{align}\label{eq:2}
G &< 2.5\,(G-J) + 10.5 \,\,\, \mathrm{for} \,\,\, G > 14.25 \, \mathrm{mag}\nonumber \\ 
G &< 2.9\,(G-J) + 9.9 \,\,\, \mathrm{for} \,\,\, G < 14.25	 \, \mathrm{mag}.
\end{align}

\subsection{Source distribution}\label{sec:3.3}
\begin{figure*}
\includegraphics[width = \hsize, keepaspectratio]{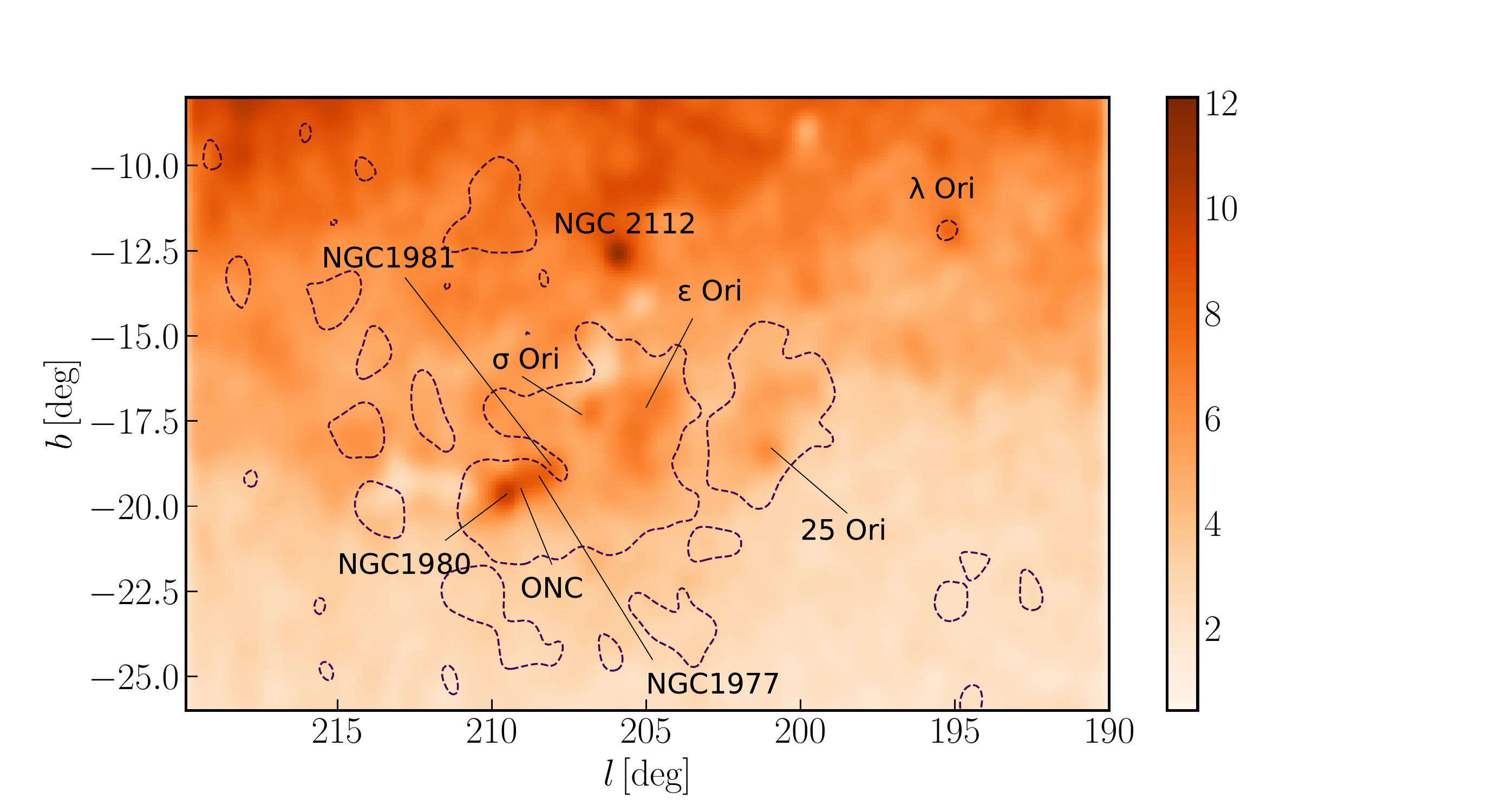}%
\caption{Normalized probability density function of the stars selected with with Eq. \eqref{eq:2} (Gaussian kernel with bandwidth = $0.03^{\circ}$). The density enhancements visible in the centre of the field (Galactic longitude between $200^{\circ}$ and $210^{\circ}$, Galactic latitude $-20^{\circ}$ and $-15^{\circ}$) are related to the TGAS density enhancements (the black dashed contours correspond to the $S = 3$ levels of the TGAS density map of Fig. \ref{fig:4}).  The peak at $(l, b) \sim (206, -12.5)$ deg corresponds to  the open cluster NGC 2112. 
}
\label{fig:7}
\end{figure*}

We choose to study the distribution in the sky of the sources selected with Eq. \eqref{eq:2} repeating the procedure explained in Sec. \ref{sec:3.1}. 
We analyse the source density using again a multivariate normal kernel, with isotropic  bandwidth =  $0.3^{\circ}$ and haversine metric.
Fig. \ref{fig:7} shows the normalized probability density function of the source distribution on the sky. The dashed contours represent the $S = 3$ levels of the TGAS density map. The density enhancements towards the centre of the field  are in the same direction as the groups shown in Fig. \ref{fig:2} and reported in Table \ref{table:1}. 
The density peak in $(l, b) \sim (206^{\circ}, -12.5^{\circ})$ is associated to the old open cluster NGC 2112 (age $\sim$ 1.8 Gyr and distance $\sim 940 \, \mathrm{pc}$, see e.g. \citet{Carraro2008} and references therein).

\begin{figure*}
\includegraphics[width = \textwidth, keepaspectratio]{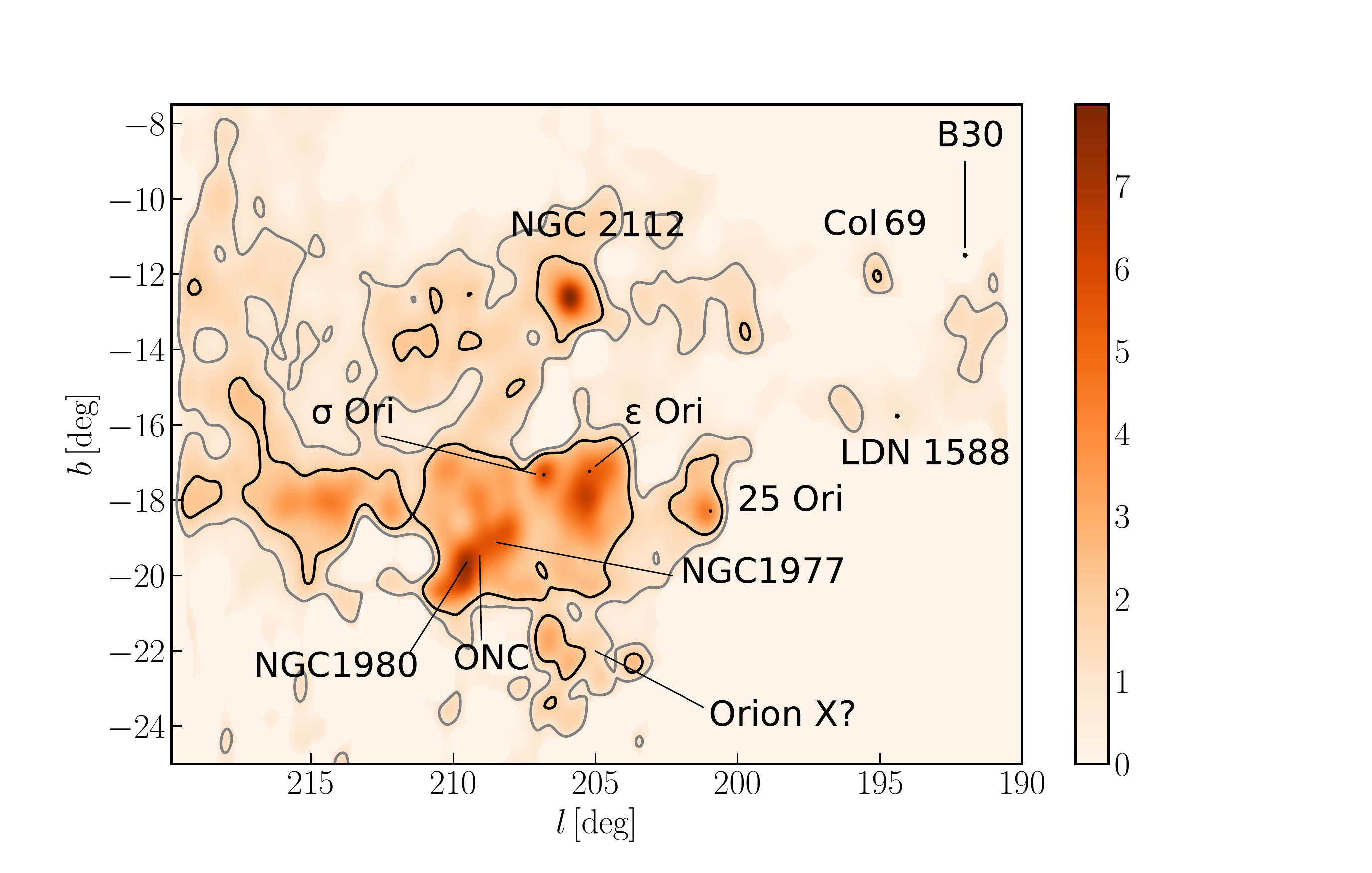}
\caption{Background subtracted kernel density estimate of the sources selected through Eq. \eqref{eq:2}. The subtraction procedure is explained in Sec. \ref{sec:3.2}. The density enhancements are highlighted by the contour levels, corresponding to $S =1$ (gray) and $S = 2$ (black).
}
\label{fig:8}
\end{figure*}

Fig. \ref{fig:8} shows $D(l, b) - \left\langle D \right\rangle$ (same notation as in Sec. \ref{sec:3.2}), and the contours represent the $S = 1$ (gray) and $S = 2$ (black) significance levels. A certain degree of contamination is present, however the groups clearly separate from the field stars.
Aside from the structures already highlighted in the TGAS map of Fig. \ref{fig:4}, some other features are visible in the KDE of Fig. \ref{fig:8}.

\begin{itemize}
\item The density enhancements towards $\lambda$ Ori include not only the central cluster (Collinder 69, $\sim (195^{\circ}, -12^{\circ}$) but also  some structures probably related to Barnard 30 ($\sim 192^{\circ}, -11.5^{\circ}$) and LDN 1588 ($\sim 194.5^{\circ}, -15.8^{\circ}$). Some small over-densities are located on the H$\alpha$ bubble to the left of LDN 1588 and they do not correspond to any  previously known group.

\item The shape of 25 Ori is elongated, and presents a northern and a southern 'extension', which are also present in the TGAS KDE of Fig. \ref{fig:4}.

\item South of $\epsilon$ Ori, a significant over-density is present, possibly related to the Orion X group, discovered by \cite{Bouy2015}.

\item Around the centre of the Orion dust ring ($\sim 214^{\circ}, -13^{\circ}$) discovered by \cite{Schlafly2015} a number of densities enhancements are present. These over-densities are visible also in the TGAS map of Fig. \ref{fig:4}, but here they are more evident.
\end{itemize}

\noindent
For the following analysis steps, we selected all the sources related to the most significant density enhancements, i.e. those within the $S = 2$ contour levels shown in Fig. \ref{fig:8}. 


\subsection{Age estimates}\label{sec:3.4}
To determine the age(s) of the population(s) we identified, we perform a Bayesian isochrone fit using a method similar to the one described in \citet{Jorgensen2005} and, more recently, in  \citet{Valls-Gabaud2014}.
\noindent
These authors used Bayesian theory to derive stellar ages based on a comparison of observed data with theoretical isochrones. Age ($t$) is one free parameter of the problem, but not the only one: the initial stellar mass ($m$) and the chemical composition ($Z$) are also considered as model parameters. We simplify the problem assuming a fixed value for $Z$. 
Using the same notation as \citet{Jorgensen2005},  the posterior probability $f(t, m)$ for the age and mass is given by:
\begin{equation}
f(t, m) = f_0(t, m)L(t, m),
\end{equation}
where $f_0(t, m)$ is the prior probability density and $L$ the likelihood function. Integrating with respect to $m$ gives the posterior probability function of the age of the star, $f(t)$.   
We assume independent Gaussian errors on all the observed quantities, with standard errors $\sigma_i$. The likelihood function is then:
\begin{equation*}
L(t, m) = \prod_{i=1}^n \left(\frac{1}{(2\pi)^{1/2}\sigma_i}\right) \times \exp{ \left(- \chi^2/2 \right)},
\end{equation*}
with: 
\begin{equation*}
\chi^2 = \sum_{i = 1}^n \left(\frac{q_i^{\mathrm{obs}}-q_i(t, m)}{\sigma_i}\right)^2,
\end{equation*}
where $n$ is the number of observed quantities, and $\mathbf{q}^{\mathrm{obs}}$ and $\mathbf{q}(t, m)$ are the vectors of observed and modelled quantities.
Following \cite{Jorgensen2005}, we write the prior as:
\begin{equation*}
f_0(t, m) = \psi(t)\xi(m),
\end{equation*}
where $\psi(t)$ is the prior on the star formation history and $\xi(m)$ is the prior on the initial mass function. We assume a flat prior on the star formation history, and a power law for the initial mass function (IMF)
\begin{equation*}
\xi(m) \propto m^{-a},
\end{equation*}
with $a = 2.7$. We choose a power law following \cite{Jorgensen2005}. We also test other IMFs, and find that the final results are not strongly dependent on the chosen IMF.
We adopt the maximum of $f(t)$ as our best estimate of the stellar age. We compute the confidence interval following the procedure explained in detail in \cite{Jorgensen2005}. It might happen that the maximum of $f(t)$ coincides exactly with one of the extreme ages considered. In this case only an upper or a lower bound to the age can be set and we call our age estimate \textit{ill defined}. On the other case, if the maximum of $f(t)$ falls within the age range considered, we call our age estimate \textit{well defined}.

\noindent 
To perform the fit we compare the observed $G$ magnitude and $G-J$ color to those predicted by the PARSEC \citep[PAdova and TRieste Stellar Evolution Code][]{Bressan2012, Chen2014, Tang2014}  library of stellar evolutionary tracks. We used isochronal tracks from $\mathrm{\log(age/yr) = 6.0}$ (1 Myr) to $\mathrm{\log(age/yr) = 8.5}$ (200 Myr), with a step of $\mathrm{\log(age/yr) = 0.01}$. We choose the range above since we are mainly interested in young (age < 20 Myr) sources. As mentioned above,  we fixed the metallicity to $Z = 0.02$, following \cite{Brown1994}. 
The isochronal tracks have an extinction correction of $A_V = 0.25 \, \mathrm{mag}$. The correction was derived computing the average  extinction towards the stars in \cite{Brown1994}.
We decided to fix the extinction to a single value mainly to keep the problem simple. Besides, we have excluded mostly of the extincted sources when we applied the criteria of Eq. \ref{eq:cuts}.
  
\noindent
We applied the fitting procedure to all the stars resulting from the selection procedure in Sec. \ref{sec:3.3}, fixing the parallax to the mean value derived in the Sec. \ref{sec:3.1}, i.e. $\varpi = 2.65 \, \mathrm{mas}$. This choice is motivated primarily by the fact that with the current data quality is not possible to precisely disentangle the spatial structure of the region. More sophisticated choices for the parallax values are described in Appendix, however, even if they lead to different single age estimates, they do not change the general conclusions of the analysis. In particular the age ranking of the groups does not change.
 
\begin{figure}
\includegraphics[width = \hsize]{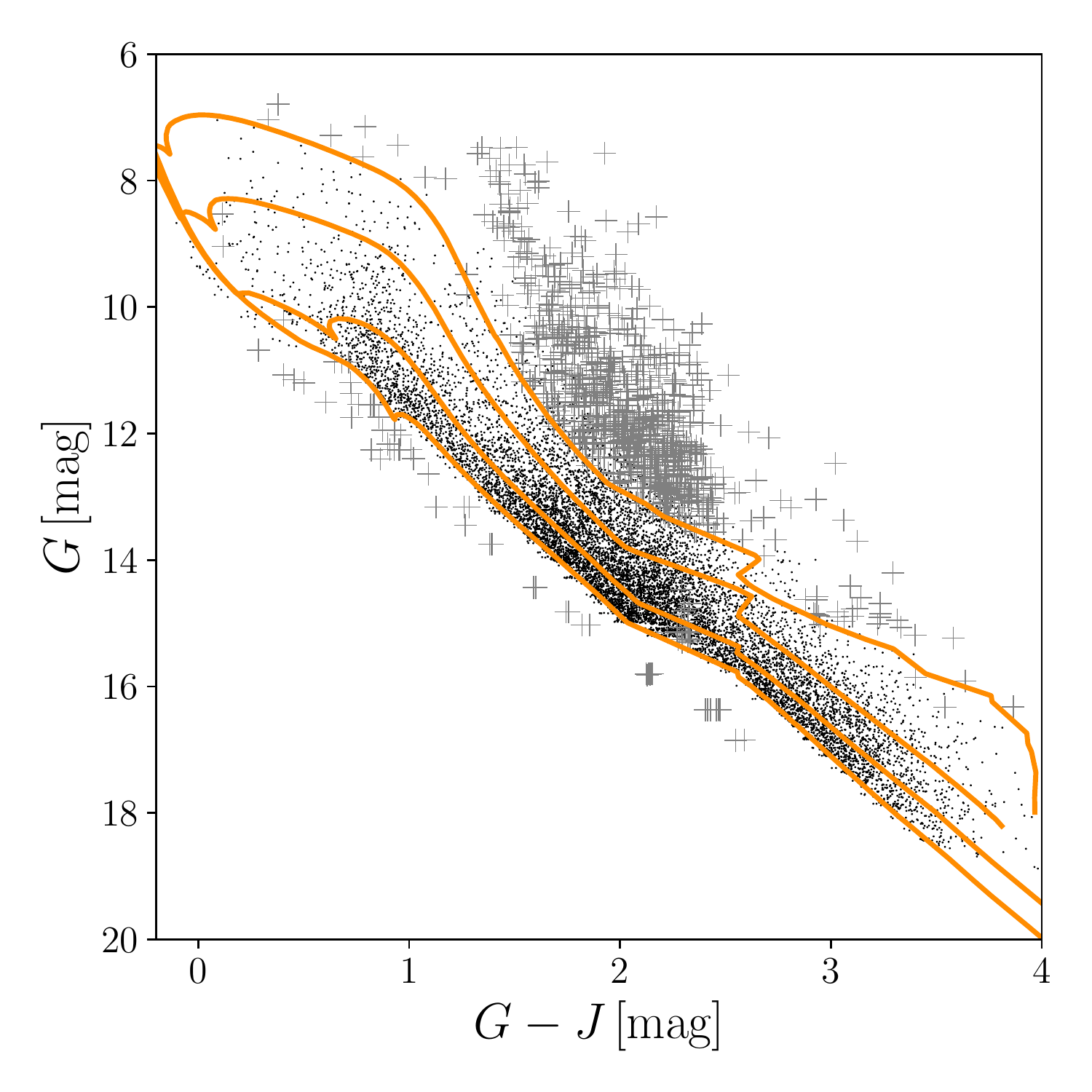}
\caption{Color magnitude diagrams of the sources with estimated age younger than 20 Myr. Black dots represent sources with well defined age estimate, gray crosses represent sources with ill-defined age estimate. The sources with ill-defined age estimates most likely belong to the Galactic disc.
The orange lines are the PARSEC isochrones at 1, 3, 10 and 20 Myr at a distance of $\sim 380 \,\mathrm{pc}$. 
}
\label{fig:6a}
\end{figure}

Fig. \ref{fig:6a} shows the color magnitude diagram of the sources with estimated age younger than 20 Myr. The gray crosses are the sources whose age is ill defined, the black dots represent the sources with well defined ages.
Noteworthy, the sources with ill-defined age consist mainly of galactic contaminants, which we could then remove from our sample.

\begin{figure*}
\includegraphics[width = \hsize]{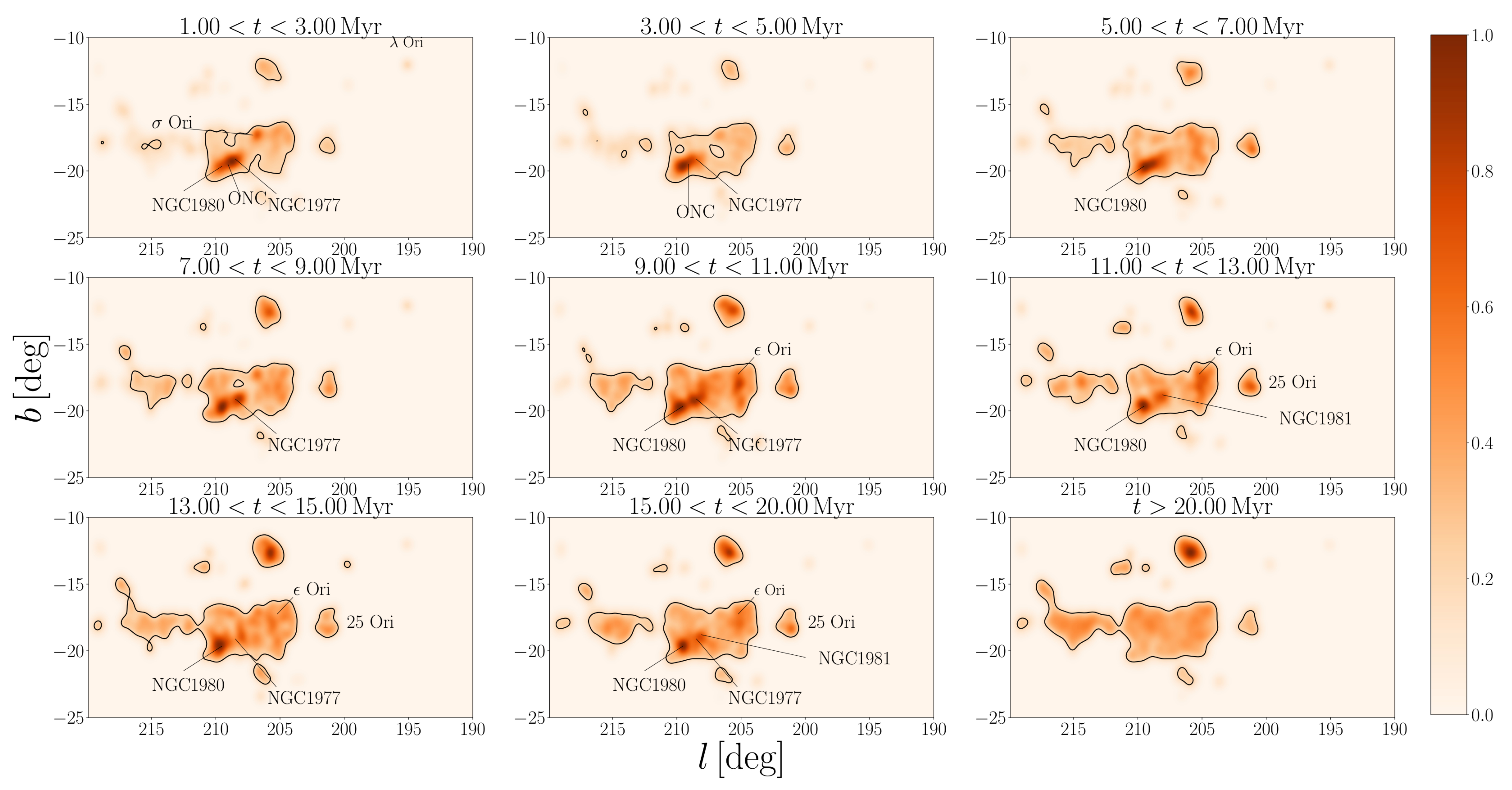}
\caption{Distribution on the sky of the sources selected in Sec. \ref{sec:3.2} for different age intervals. The ages are computed using the isochrone fitting procedure described in Sec. \ref{sec:3.4}. The contours represent the 0.05 density level and are shown only for visualization purposes. Note how the position of the density enhancements changes depending on the age. The first eight panels show stars with estimated ages < 20 Myr, while the last one shows older sources. The young stars are not coeval, in particular the age distribution shows a gradient, going from 25 Ori and $\epsilon$ Ori towards the ONC and NGC 1980. The last panel shows the field stars, whose estimated age is older than 20 Myr.}
\label{fig:9}
\end{figure*}

Fig. \ref{fig:9} shows the density (obtained with a Gaussian kernel, with bandwidth = $0.05^{\circ}$) of the source sky distribution as a function of their age, $t$. The densities are normalized to their individual maximum, so that their color scale is the same. The coordinates of the density enhancements change with time. This means that the groups we identified have different relative ages:
\begin{itemize}
\item \textit{$\sigma$ Ori.}
The peak associated $\sigma$ Ori ($(l, b) = (207, -17.5)$ deg) is in the first panel ($1 < t < 3$ Myr), and some residuals are present also in the second panel ($3 < t < 5$ Myr) and in the fourth ($7 < t < 9 \, \mathrm{Myr}$). 
\cite{Hernandez2007}, \cite{Sherry2008}, and \cite{Zapatero2002}  all  estimate an age  of  2-
4 Myr, which is compatible with what we find. Instead, \cite{Bell2013} puts the cluster at 6 Myr. 

\item \textit{25 Ori.}
The 25 Ori group ($(l, b) = (20.1, -18.3)$ deg) appears in the third panel ($5 < t < 7 $ Myr), peaks in the sixth panel ($9 < t < 11 $ Myr) and then fades away. \cite{Briceno2007} found that the age of 25 Ori is $\sim 7-10$ Myr. Our age estimate is slightly older, but still fits the picture of 25 Ori being the oldest group in the region. 

\item{\textit{Belt population.}}
The population towards $\epsilon$ Ori ($(l, b) \sim (205.2, -17.2)$ deg) becomes prominent for $t >  9$ Myr. Here, \cite{Kubiak2016} estimated the age to be older than $\sim 5$ Myr, without any other constraint.

\item{\textit{ONC, NGC 1980, NGC 1981, and NGC 1977.}}
The over-densities associated with NGC 1980, NGC 1981, NGC 1977 and the ONC ( centred in $(l, b) \sim (209, -19.5)$ deg)  are very prominent until the eighth  panel of Fig. \ref{fig:9}. In this last case it is difficult to disentangle exactly which group is younger, especially because the underlying data point distribution is smoothed by the Kernel.
The density enhancement in the first panel ($1 < t < 3 \, \mathrm{Myr}$) is most likely related to the ONC and L1641 \citep{Reggiani2011, DaRio2014, DaRio2016}.
The density enhancement associated with NGC 1977 peaks in the same age ranges  ($7 < t < 9 \, \mathrm{Myr}$) as the one associated with NGC 1980, which however remains visible until later ages ($15 < t < 20 \, \mathrm{Myr}$) and fades away only for $t > 20 \, \mathrm{Myr}$. Finally, the density enhancement associated with NGC 1981 does not clearly stand out in any panel, excluding perhaps the ones with age $11 < t < 13 \, \mathrm{Myr}$ and $13 < t < 15 \, \mathrm{Myr}$. 
An interesting feature of the maps is the fact that the shape and position of the density enhancements related to NGC 1980 change with time. In particular, for early ages only one peak is present, while from $\sim$7 Myr two peaks are visible. This is a further confirmation that the density enhancements in 
the first three age panels include L1641 and the ONC, which are indeed younger than the other groups. 
\cite{Bouy2014} derived an age $\sim 5-10$ Myr for NGC 1980 and NGC 1981.

\end{itemize}
The last panel shows the stars with estimated ages > 20 Myr. The source distribution is uniform. These are field stars, with estimated ages ranging from 20 to 200 Myr. 

Our fitting procedure does not take into account the presence of unresolved binaries among our data. Since the sample includes pre-main sequence stars, the binary population could be mistaken for a younger population at the same distance. For example, the binary counterpart of a population with age $t \sim 12 $ Myr falls in the same locus of the $G-J$ vs $G$ color magnitude diagram as a population with age $t \sim 7$  Myr. This means that the fit could mistake the unresolved binaries for a younger population, therefore the interpretation of Fig. \ref{fig:9} requires some care. 
Another caveat is related to the definition of the \textit{Gaia G}  band in the PARSEC libraries. Indeed, the nominal \textit{Gaia G}  passband \citep{Jordi2010} implemented in the PARSEC libraries  is different from the actual one \citep[cfr. ][]{Carrasco2016}. This affects the values of \textit{G} and \textit{G-J} predicted by the PARSEC libraries and therefore our absolute age estimates, but does not influence the age ordering. 
The same can be said for the extinction. Choosing a different (constant) extinction value shifts the isochronal tracks, and therefore the estimated age is different, but does not modify the age ranking.
In conclusion, the age ranking we obtain is robust, and, even with all the aforementioned cautions, Fig. \ref{fig:9} shows the potential of producing age maps for the Orion region.

\section{Orion in Pan-STARRS1} 
To confirm the age ordering we obtain with \textit{Gaia} DR1,  we apply the analysis described in Sec. \ref{Sec:3} to the recently published Pan-STARRS1 photometric catalogue \citep{Chambers2016,Magnier2016}.

Pan-STARRS1 has carried out a set of distinct synoptic imaging sky surveys including the $3\pi$ Steradian Survey and the Medium Deep Survey in 5 bands ($grizy$). The mean 5$\sigma$ point source limiting sensitivities in the stacked 3$\pi$ Steradian Survey in $grizy$ are (23.3, 23.2, 23.1, 22.3, 21.4) magnitudes respectively. For stars fainter than $r \sim 12 \, \mathrm{mag}$, Pan-STARRS1 and \textit{Gaia} DR1 photometric accuracies are comparable. Stars brighter than $r \sim 12 \, \mathrm{mag}$ have large photometric errors in the PanSTARRS filters, therefore we decide to exclude them from our sample. 
We consider the same field defined in Eq. \eqref{eq:1} and we perform a cross-match of the sources with \textit{Gaia} DR1 and 2MASS, using a cross-match radius of 1''. We do not account for proper motions, since the mean epoch of the Pan-STARRS1 observations goes from 2008 to 2014 for the cross-matched stars and therefore the cross-match radius is larger than the distance covered in the sky by any star moving with an average proper motion of a few $\mathrm{mas \, yr^{-1}}$.
We obtain $N = 88 \, 607$ cross-matched sources, and we analyse this sample with the same procedure explained in Sec. 3. Briefly, we first exclude the bulk of the field stars making a cut in the $r-i$ vs. $r$ color-magnitude diagram:
\begin{equation}
r <  5 \times (r-i)+ 12 \, \mathrm{mag}.
\end{equation}
Then we perform the same $JHK$ photometric selection as in Eq. \ref{eq:2}, and we study the on-sky distribution of the sources. We find some density enhancements, corresponding to those already investigated with the \textit{Gaia} DR1 only. We then smooth
the data point distribution in Galactic coordinates using a multivariate Gaussian kernel with bandwidth $0.3^{\circ}$. 
We select all the sources within the $S = 2$ density levels and we estimate the single stellar ages with the same Bayesian fitting procedure described above. 
In this case however we do not use the \textit{Gaia} and 2MASS photometry, but the $r$ and $i$ Pan-STARRS1 bands.  

\begin{figure*}
\includegraphics[width = \hsize]{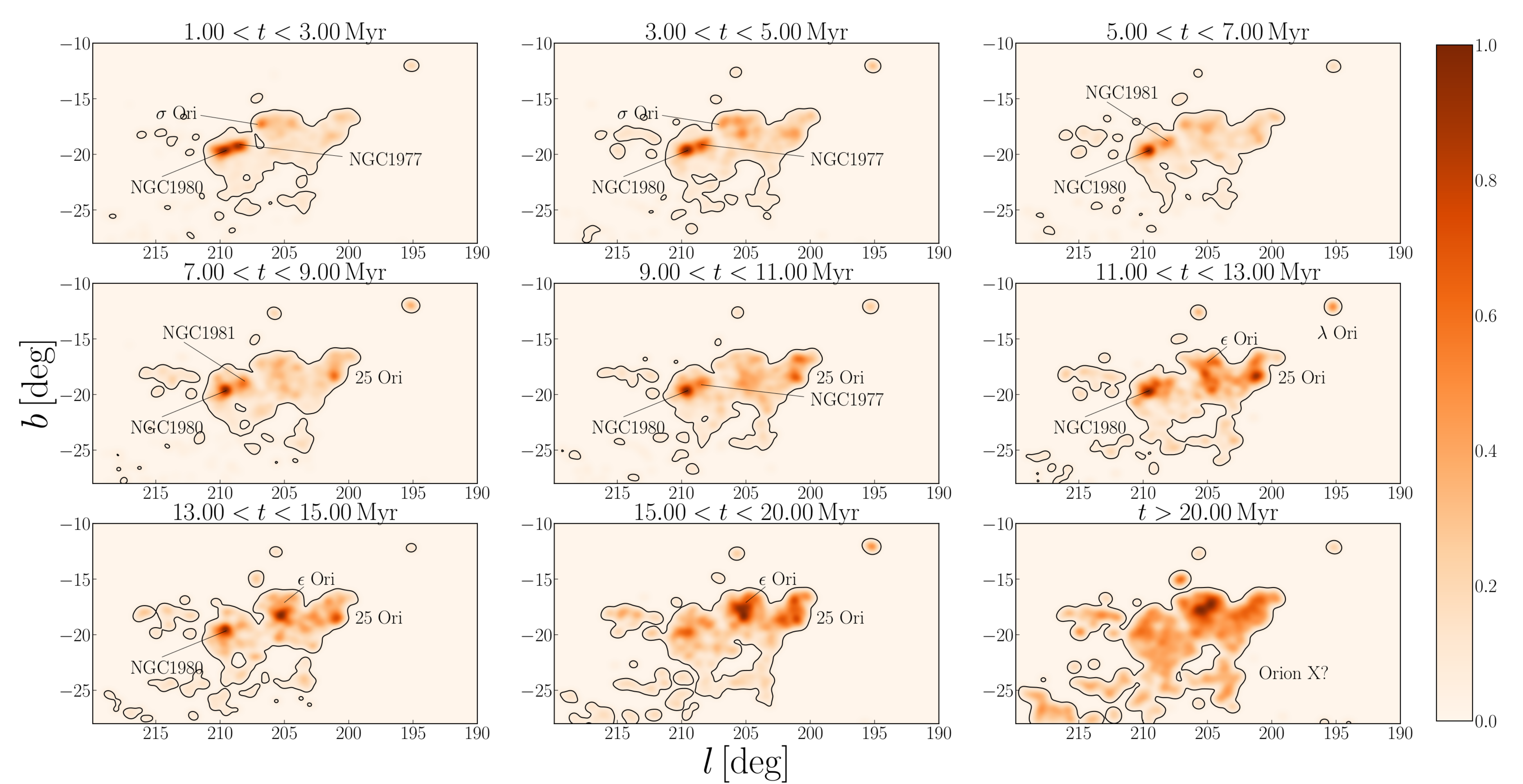}
\caption{Same as Fig. \ref{fig:7} but using the Pan-STARRS1 $r$ and $i$ band to derive ages. The contours represent the 0.05 density levels and are shown only for visualization purposes.}
\label{fig:10}
\end{figure*}

Fig. \ref{fig:10} shows the on-sky distribution of the sources with similar ages. The age intervals used are the same as in Fig. \ref{fig:9}. The density enhancements corresponding to known groups are visible. Moreover, by comparing Figs. \ref{fig:9} and \ref{fig:10}, one can immediately notice that the same groups appear in the same age intervals except for  the $\epsilon$ Ori group, that appears slightly older than with \textit{Gaia} DR1 photometry. Indeed the $\epsilon$ Ori density enhancement peaks in $15 < t < 20 \, \mathrm{Myr}$ with PanSTARRS photometry, while it is spread between $11 < t < 20 \, \mathrm{Myr}$ with \textit{Gaia} DR1. 
Another interesting feature of the Pan-STARRS1 age maps are the density enhancements below $\epsilon$ Ori. These structures appear prominently in the oldest age panels, and might be related to the Orion X population \citep{Bouy2015}.

These results strengthen our confidence in the age estimates obtained with \textit{Gaia} photometry, in particular regarding the age ordering. 

\section{Discussion}
The present analysis confirms the presence of a large and diffuse young population towards Orion, whose average distance is  $d \sim 380 \, \mathrm{pc}$.
The ages determined in Sec. \ref{sec:3.4} show that the groups are young (age < 20 Myr) and not coeval. The age ranking determined using \textit{Gaia} and 2MASS photometry (Fig. \ref{fig:7})  is  consistent with that determined using Pan-STARRS1 (Fig. \ref{fig:10}). 

\noindent
Figs. \ref{fig:8}, \ref{fig:9}, and \ref{fig:10}  show some important features, which can potentially give new insights on our understanding of the Orion region.

\textit{The Orion dust ring.} As already mentioned in Sec. 3.3, a number of over-densities are present towards the Orion dust ring discovered by \cite{Schlafly2015}. The age analysis is not conclusive since many over-densities are not within $S = 2$. Unfortunately, there are no proper motions and/or parallaxes available for these sources (nor in \textit{Gaia} DR1 nor in other surveys), and their distribution in the color-magnitude diagram is not very informative. Additional clues about their origin will be hopefully provided by \textit{Gaia} DR2.

\begin{figure*}
\includegraphics[width = 0.5\hsize]{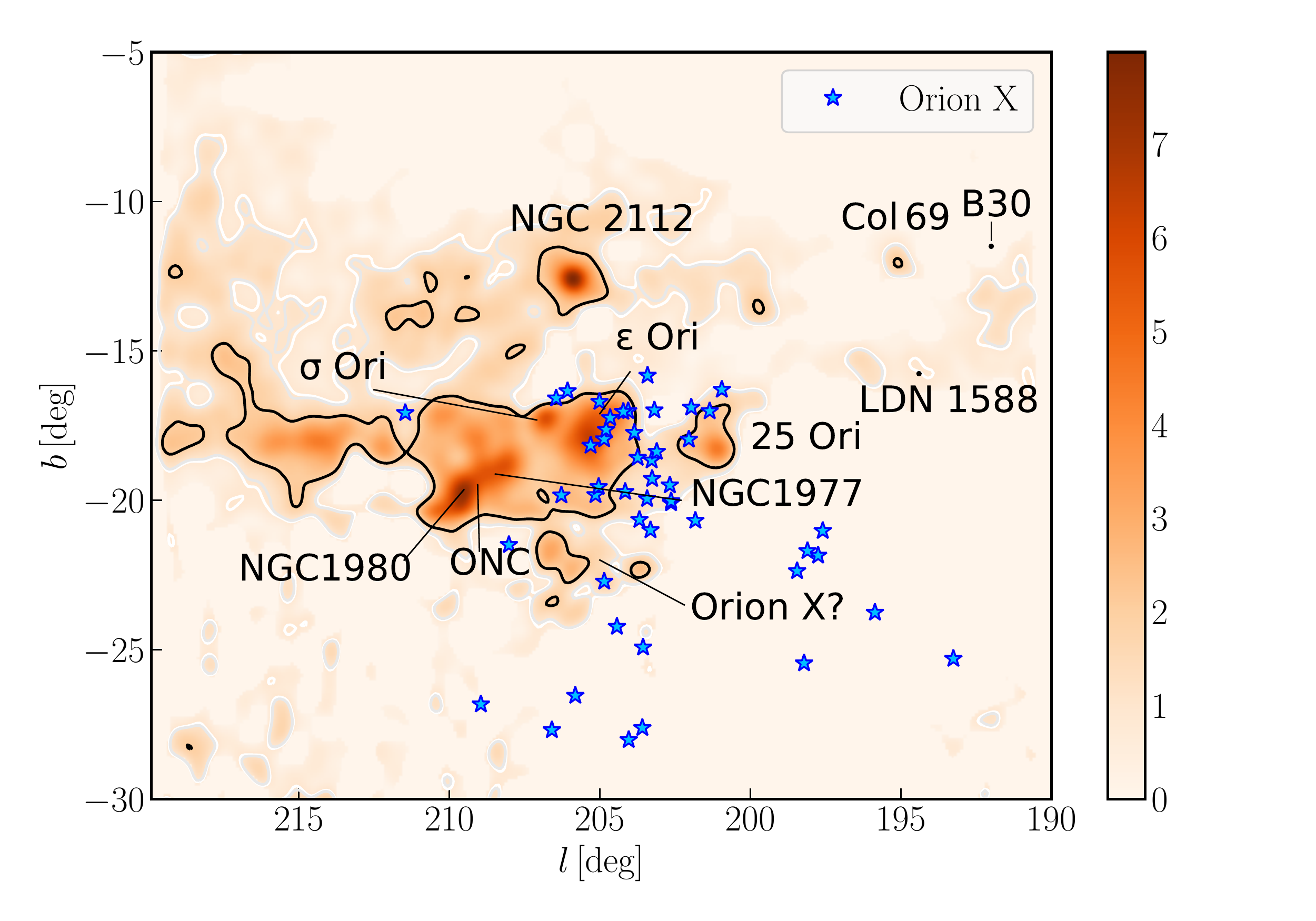} 
\vspace{0.1cm}
\includegraphics[width = 0.5\hsize]{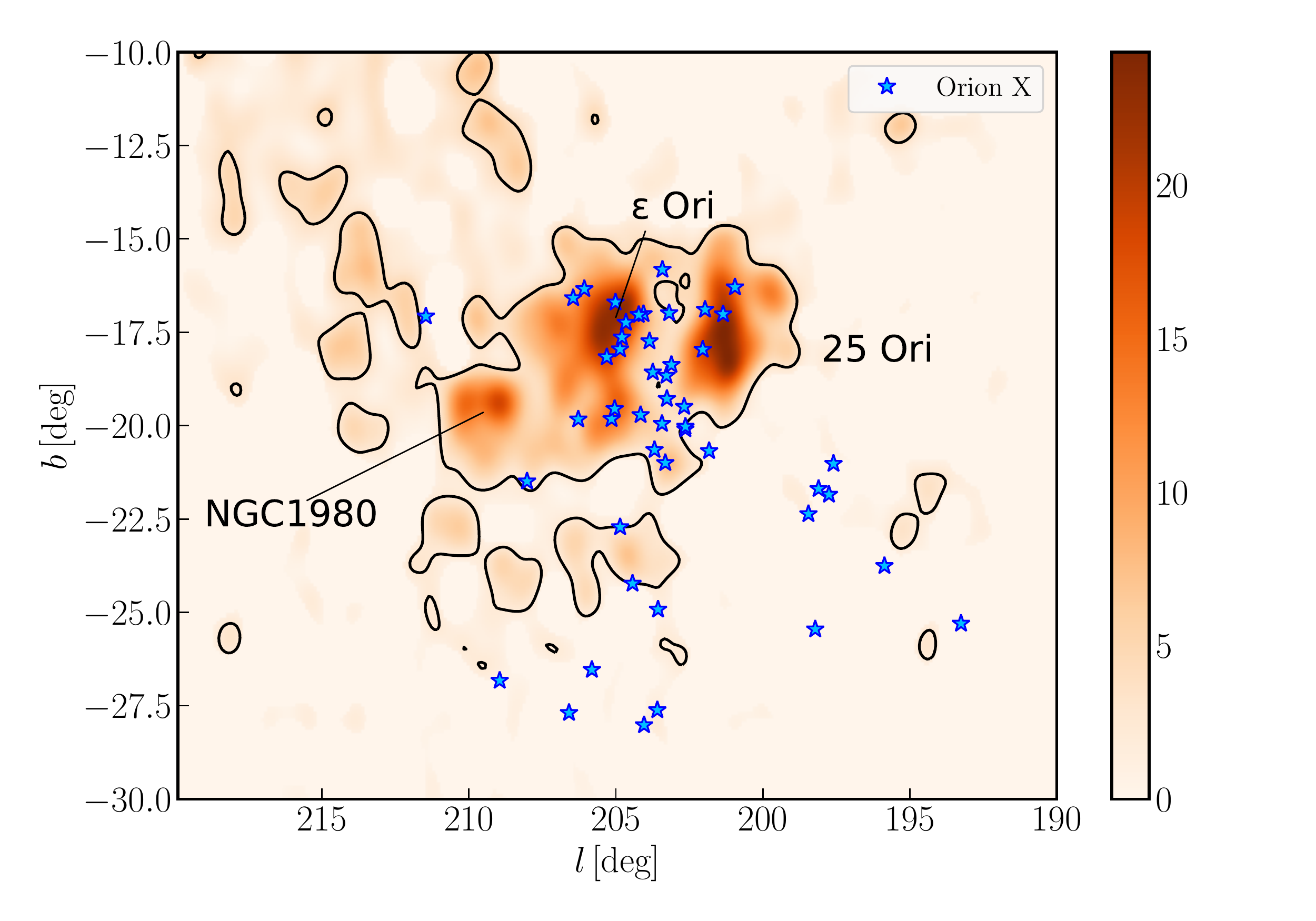} 
\caption{Left: The Orion X candidate members from \cite{Bouy2015} are plotted over the kernel density estimation of Fig. \ref{fig:8} as blue stars. Right: The Orion X candidate members are plotted over the kernel density estimation of the TGAS sources with $3 < \varpi < 7 \, \mathrm{mas}$.}
\label{fig:11}
\end{figure*}

\textit{The Orion Blue-stream.} \cite{Bouy2015} studied the 3D spatial density of OB stars in the Solar neighbourhood and found three large stream-like structures, one of which is located towards $l \sim 200^{\circ}$ in the Orion constellation (Orion X). Fig. \ref{fig:11} shows the position of the candidate members of the Orion X group as blue stars. Even though the candidate member centre looks slightly shifted with respect to the density enhancements shown in the map, it is difficult to argue that these stars are not related to the young population we analysed in this study. \cite{Bouy2015} report that the parallax distribution of the Orion X sources goes from $\varpi \sim 3 \mathrm{mas}$ to $\varpi \sim 6 \, \mathrm{mas}$ ($150 < d < 300 \, \mathrm{pc}$), which indicates that Orion X is  in the foreground of the Orion complex. \cite{Bouy2015}  also propose that the newly discovered complex could be older than Orion OB1 and therefore constitute the front edge of a stream of star formation propagating further away from the Sun.

\begin{figure}
\includegraphics[width = \hsize]{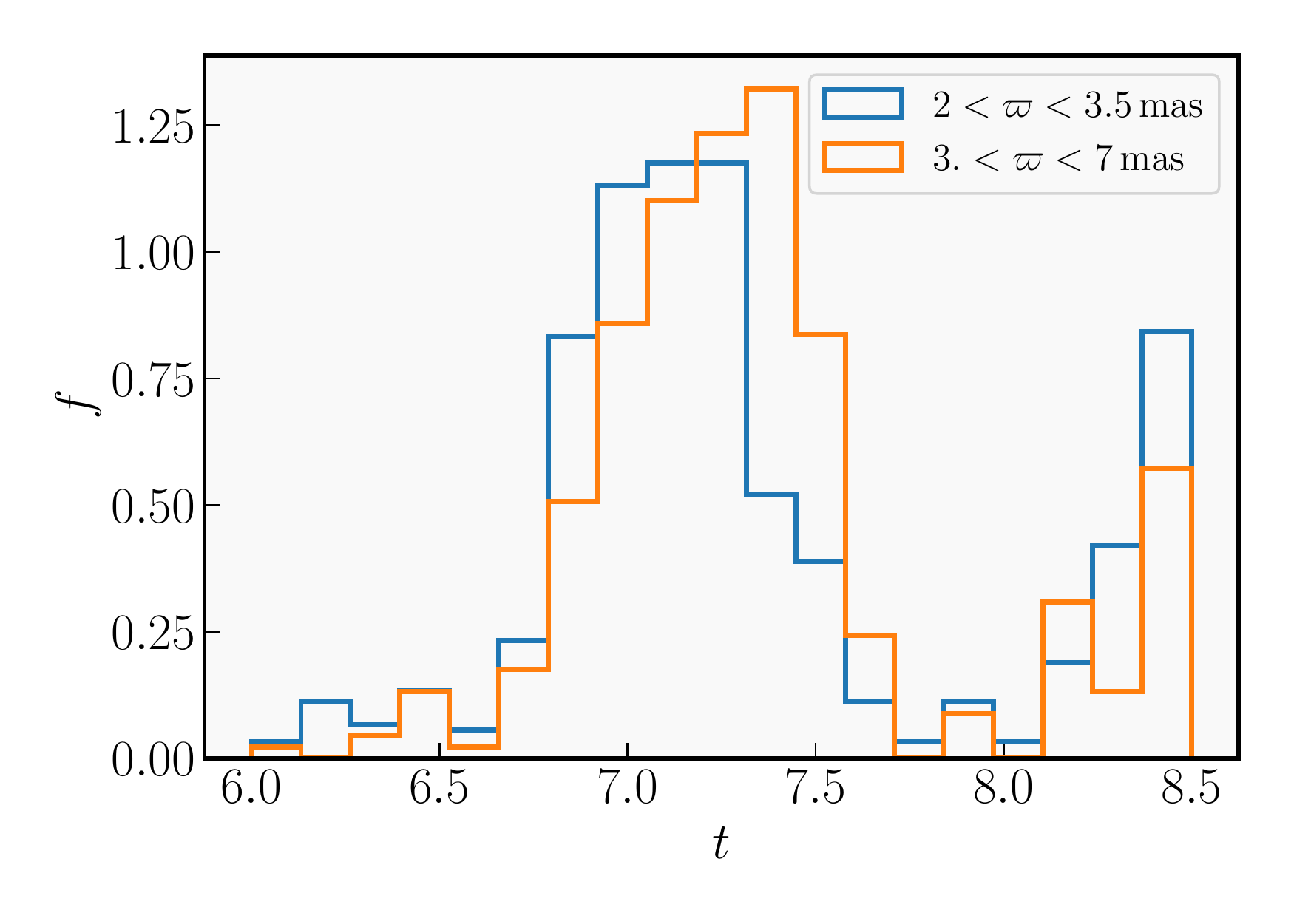}
\caption{Age distribution of the TGAS sources with $2 < \varpi < 3.5 \, \mathrm{mas}$ (blue) and $3. < \varpi < 7 \, \mathrm{mas}$ (orange). The median of the distributions is respectively $t = 7.19 \log(age/yr)$ ($\sim 15 \, \mathrm{Myr}$) and $t = 7.27  \log(age/yr)$ ($\sim 19 \, \mathrm{Myr}$).}
\label{fig:11b}
\end{figure}

To test this  scenario we proceeded as follows.
First we complemented the bright end of TGAS with Hipparcos data, then we selected the stars using the proper motion criterion of Eq. \eqref{eq:3} and with $3 <\varpi < 7 \,\mathrm{mas}$. In this way we restricted our sample to the  stars probably kinematically related to the Orion OB association, but on average closer to the Sun.  The density of the distribution of theses sources in the sky is shown in Fig. \ref{fig:11}, together with the Orion X candidate members. We selected the sources within the S = 2 levels (with S defined in Section 3), and we used the Bayesian isochronal fitting procedure to estimate the age of this population. 
Note that out of the 48 Orion X candidate members listed in \cite{Bouy2015}, only 22 are included in TGAS (the others are probably too bright).
To perform the isochronal fit, we could actually use the measured parallax, instead of one single value. The age distribution for the foreground sources is shown in Fig. \ref{fig:11b} (orange histogram). As a comparison, the age distribution of the sources within the density enhancements and with $2 < \varpi < 3.5 \, \mathrm{mas}$ is also shown (blue histogram). On average, the foreground population looks older, which is consistent with the picture that \cite{Bouy2015} proposed. There are however two caveats: 
\begin{itemize}
\item the age distributions are broad;
\item  the parallax errors are large and dominate the age estimate. 
\end{itemize}
With future \textit{Gaia} releases we will be able to further study the Orion X population and more precisely characterize it.

\textit{25 Ori.} As pointed out in Sec. 3.3 the 25 Ori group presents a northern extension ($\sim 200^{\circ},-17^{\circ}$) visible in the TGAS, Gaia DR1 and Pan-STARRS1 density maps. The northern extension parallax is only slightly larger than that of the 25 Ori group, and the age analysis suggests that the groups are coeval. With a different approach, \cite{Lombardi2017} find evidence of the same kind of structure (see their Fig. 15). \textit{Gaia DR2} will be fundamental in discerning the properties of this new substructure of the 25 Ori group.

\textit{The $\lambda$ Ori group.} In Sec. 3.3 we pointed out some over-densities located on the H$\alpha$ bubble surrounding $\lambda$ Ori, which are not related to known groups (to our knowledge). We further investigated the stars belonging to these over-densities, however there are no parallaxes nor proper motions available for these sources and it is difficult to draw firm conclusions from the photometry only (also combining \textit{Gaia} DR1 and Pan-STARRS1). In this case as well, we have to conclude that hopefully \textit{Gaia} DR2 will clarify if this groups are real or not.

\textit{NGC 1980 and the ONC.} One of the most interesting features of the maps of Fig. \ref{fig:9} and Fig. \ref{fig:10} is the prominent density enhancement towards NGC 1980, NGC 1977 and the ONC. The density enhancement is not concentrated in only one panel, but persists in all of them and disappears in the last one. This can be explained in at least two ways:
\begin{itemize} 
\item there are multiple populations at roughly the same distance, with different ages; 
\item there is only one population with a single age, however its spread along the line of sight is so large that using only one parallax value for the fit is not accurate enough.
\end{itemize}
Both explanations have supporters. 
\cite{Alves2012} suggested that NGC 1980 is not directly related to the ONC, i.e. they are not the same population emerging from its parental cloud but are instead distinct overlapping populations. On the other hand, based on the fact that the kinematic properties of NGC 1980 are indistinguishable from those of the rest of the population at the same position in the sky, \cite{DaRio2016} argued that NGC 1980 simply represents the older tail of the age distribution around the ONC, in the context of an extended star formation event. Using isochronal ages, \citet{Fang2017} find that the foreground population has a median age of 1-2 Myr, which is similar to that of the other young stars in Orion A. Furthermore they confirm that the kinematics of the foreground population is similar to that of the molecular clouds and of other young stars in the region. They therefore argue against the presence of a large foreground cluster in front of Orion A. \cite{Kounkel2017b} estimate that the age of NGC 1980 is $\sim 3 \, \mathrm{Myr}$, which is comparable with the study by \citet{Fang2017}, however they are not able to confirm or disprove whether NGC 1980 is in the foreground on the ONC. 
Finally, \citet{Beccari2017} discovered three well-separated pre-main sequences in the $r-i$ vs $r$ color-magnitude diagram obtained with the data of the wide field optical camera OmegaCAM on the VLT Survey Telescope (VST) in a region around the ONC. These sequences can be explained as a population of unresolved binaries or as three populations with different ages. The populations studied by Beccari et al. are unlikely to be related to NGC 1980, however, if confirmed, they would constitute an example of non-coeval populations in the same cluster.
Fig. \ref{fig:9} shows that the group corresponding to NGC 1980 is well defined not only at very young ages ($1 < t < 3 \, \mathrm{Myr}$), but at least until $t \sim 15 \mathrm{Myr}$. We will discuss below the influence that unresolved binaries have on our age determination (indeed our fit does not account for them), the main point being that unresolved binaries influence the youngest age intervals, not the oldest. This would point towards the actual existence of two populations, the first related to the ONC, the second to the \cite{Alves2012} foreground population.

In conclusion, the ages of the stellar populations towards Orion show a gradient, which goes from 25 Ori and $\epsilon$ Ori towards the ONC and the Orion A and B clouds.
The age gradient is also associated to a parallax gradient: indeed the older population towards 25 Ori and $\epsilon$ Ori is also closer to the Sun than the younger one towards the ONC (see also Fig. \ref{fig:13}). \textit{Gaia} DR2 will provide distances to the individual stars of each different group, and we will therefore be able to obtain also more precise ages for them.

To study whether or not the parallax gradient influences the age determination, we performed the same Bayesian isochrone fit changing each star's parallax according to its position, following  Fig. \ref{fig:13}.
We also performed the analysis including a uniform prior on the parallax distribution, and then marginalizing over the parallax. 
In both cases, the estimated ages for the single groups have some small variations, however our conclusions do not significantly change.
 
To test how our  result depend on the set of isochrones we chose, we performed the fit again, using the MESA Isochrones and Stellar Tracks \citep[MIST][]{Mist0, Mist1}. We fixed the metallicity to $\mathrm{Z_{\odot}}$ and we applied the usual extinction correction of $A_V = 0.25 \, \mathrm{mag}$. Whilst in this case the single ages are in general estimated to be younger than with the PARSEC models (e.g., the 25 Ori group peaks between  $9 < t < 11 \, \mathrm{Myr}$), the age ordering does not change significantly.
 
Finally, we studied the distribution in the sky of the coeval sources fainter than 14 magnitudes. In this magnitude range we can remove the sources that are most likely Galactic contaminants. We found again the same groups and the same age ordering.

As mentioned above and in Sec. \ref{sec:3.4}, the unresolved binary sequence could stand out as a separate, seemingly younger population, which would add further complications to the age determination of the group. The Bayesian fitting procedure does not take into account the presence of unresolved binaries. The net effect of this is that the unresolved binaries population is mistaken for a younger population. 
For example, the difference in magnitudes between the 5-7 Myr and the 13-20 Myr isochrones correspond almost exactly to the 0.75 mag separating the primary sequence from the unresolved binary sequence. This is a major cause of age spread and it could greatly affect our age estimates, thus it appears even clearer that great care needs to be used when analysing them.
On the other hand however, binary should affect all populations in the same way. This further support the robustness of our relative age estimates.

\begin{figure*}
\includegraphics[width = \hsize]{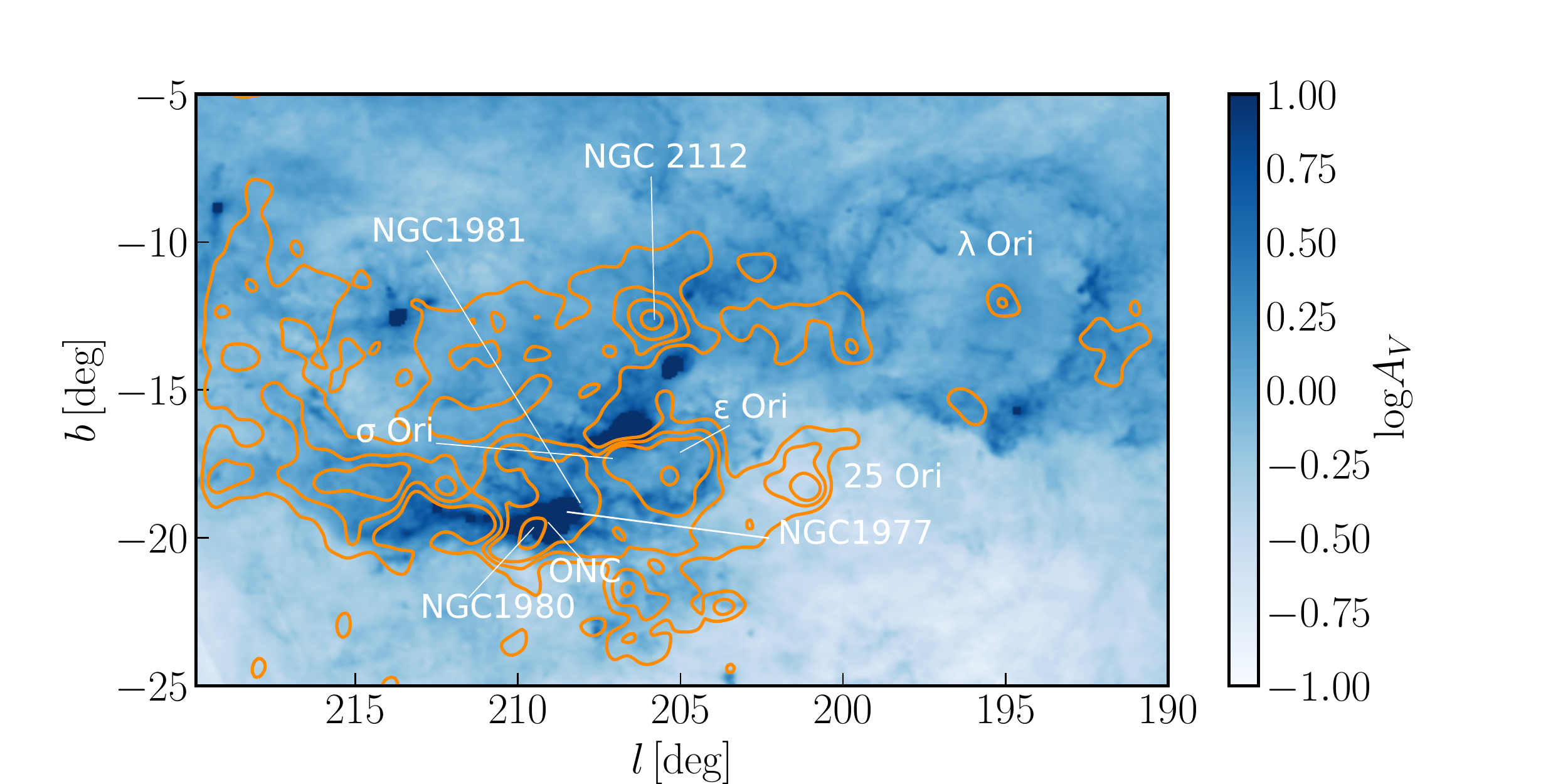}
\caption{Planck extinction map of the Orion field \citep{Planck2014}. The contour levels represent the $S = 1,2,3$ and 6 levels of the density distribution shown in Fig. \ref{fig:7}.}
\label{fig:16}
\end{figure*}

Another intriguing problem is related to the relation between the the density enhancements, the diffusely distributed massive stars, and the gas distribution. 
Fig. \ref{fig:16} shows the $S = 3, 6$ and $9$ contour levels of the over-densities on top of an extinction map obtained with \textit{Planck} data \citep{Planck2014} probing the dark clouds. The older group 25 Ori is located far away from the gas, while the younger groups of $\lambda$ Ori, $\sigma$ Ori, $\epsilon$ Ori, NGC 1977 and NGC 1980 closely follow the clouds. Orion A and B are behind the density enhancements.
The three dimensional structure of the region is still unclear, and the current data accuracy is not yet good enough to draw definite conclusions, especially at the distance and direction of Orion. The data quality however will improve  in future \textit{Gaia} releases, and likewise our understanding of the region. In particular, precise parallaxes, proper motions, and radial velocities will allow us to address directly the recent discovery that the Orion clouds might be part of an ancient dust ring \citep{Schlafly2015}, the blue streams scenario proposed by \cite{Bouy2015}, and the complex nested shell picture unveiled by \cite{Ochsendorf2015}.


\section{Conclusions}
In this paper we made use of  \textit{Gaia} DR1 \citep{Brown2016, Prusti2016, vanLeeuwen2017} to study the stellar populations towards Orion. Our results are as follows:

\begin{itemize}
\item Using TGAS \citep{Michalik2015, Lindegren2016} we found evidence for the presence of a young population, at a parallax $\varpi \sim 2.65 \, \mathrm{mas}$  ($d \sim 377 \, \mathrm{pc}$), loosely distributed around some known clusters: 25 Ori, $\epsilon$ Ori and $\sigma$ Ori, and  NGC 1980 and ONC. The stars belonging to these groupings define a sequence in all the color magnitude diagrams constructed by combining \textit{Gaia} DR1 and 2MASS photometry.
\item We considered the entire \textit{Gaia} DR1, again realizing color magnitude diagrams combining \textit{Gaia} and 2MASS photometry for the entire field. Well visible between $G = 14 \, \mathrm{mag}$  and $G = 18 \, \mathrm{mag}$, we found the low mass counterpart of the sources isolated with TGAS. 
\item After a preliminary selection to exclude field stars, we studied the distribution in the sky of the sources belonging to this sequence using a Kernel Density Estimation (KDE). We found  density enhancements in the sky distribution comparable to those in the TGAS sample.
\item We estimated the ages of the sources within the density enhancements, using  a Bayesian isochrone fitting procedure described in detail in \cite{Jorgensen2005}. We  assumed all the stars to be at the same parallax, $\varpi = 2.65$. We found that the groupings have different ages. In particular, there is an age gradient going from 25 Ori (13-15 Myr) to the ONC (1 Myr).
 \item To consolidate our findings, we repeated the fitting procedure using the sources in common with Pan-STARRS1 \citep{Chambers2016, Magnier2016} $r$ and $i$ filters, finding the same age ordering as with \textit{Gaia} DR1. 
 
\item We studied the distribution in the sky of the groups we found. In particular:
\begin{enumerate}
\item The 25 Ori cluster presents a northern extension, reported also by \cite{Lombardi2017}.
\item Some of the density enhancements towards the $\lambda$ Ori complex are related to known clusters (Col 69, B30, and $\mathrm{LDN \, 1588}$), but some other over-densities on the left of the ring are new. Unfortunately it was not possible to investigate them further since we  have neither precise proper motions, nor parallaxes.
\item Some over-densities are also present within the Orion dust ring discovered by \cite{Schlafly2015}, and they might be related to the star formation process out of which the ring was formed. In this case as well however, more data are needed to confirm our speculations.
\item The Orion X candidate members \citep{Bouy2015} are related to some of the density enhancements shown in Fig. \ref{fig:9}. We studied the sky and age distribution of the TGAS sources with proper motions as in Eq. \eqref{eq:3} and parallax $3 < \varpi < 7 \, \mathrm{mas}$, and we found that the stars with $2 < \varpi < 3.5 \, \mathrm{mas}$ are on average younger than those with $3 < \varpi < 7 \, \mathrm{mas}$.
\end{enumerate}

\item We discussed the implications of the age ranking we obtained. 
We found that the estimated ages towards the NGC 1980 cluster span a broad range of values. This can either be due to the presence of two populations coming from two different episodes of star formation or to a large spread along the line of sight of the same population. Some confusion might arise also from the presence of unresolved binaries, which are not modelled in the fit, and usually stand out as a younger population. We related our findings to previous works by \cite{Bouy2014, DaRio2016} and \cite{Fang2017}. 
\item  Finally, we link the stellar groups to the gas and dust features in Orion, albeit in a qualitative and preliminary fashion. Future \textit{Gaia} releases will allow to address these questions in unparalleled detail.
\end{itemize}

\begin{acknowledgements}
We are thankful to the anonymous referee, for comments that greatly improved the manuscript.
This project was developed in part at the 2016 NYC Gaia Sprint, hosted by the Center for Computational Astrophysics at the Simons Foundation in New York City, and at the 2017 Heidelberg Gaia Sprint, hosted by the Max-Planck-Institut für Astronomie, Heidelberg. \\
This work has made use of data from the European Space Agency (ESA)
mission {\it Gaia} (\url{https://www.cosmos.esa.int/gaia}), processed by
the {\it Gaia} Data Processing and Analysis Consortium (DPAC,
\url{https://www.cosmos.esa.int/web/gaia/dpac/consortium}). Funding
for the DPAC has been provided by national institutions, in particular
the institutions participating in the {\it Gaia} Multilateral Agreement.
This publication has made use of data products from the Two Micron All Sky Survey, which is a joint project of the University of Massachusetts and the Infrared Processing and Analysis Center/California Institute of Technology, funded by the National Aeronautics and Space Administration and the National Science Foundation. 
The Pan-STARRS1 Surveys (PS1) and the PS1 public science archive have been made possible through contributions by the Institute for Astronomy, the University of Hawaii, the Pan-STARRS Project Office, the Max-Planck Society and its participating institutes, the Max Planck Institute for Astronomy, Heidelberg and the Max Planck Institute for Extraterrestrial Physics, Garching, The Johns Hopkins University, Durham University, the University of Edinburgh, the Queen's University Belfast, the Harvard-Smithsonian Center for Astrophysics, the Las Cumbres Observatory Global Telescope Network Incorporated, the National Central University of Taiwan, the Space Telescope Science Institute, the National Aeronautics and Space Administration under Grant No. NNX08AR22G issued through the Planetary Science Division of the NASA Science Mission Directorate, the National Science Foundation Grant No. AST-1238877, the University of Maryland, Eotvos Lorand University (ELTE), the Los Alamos National Laboratory, and the Gordon and Betty Moore Foundation.
\\
This research made use of Astropy, a community-developed core Python package for Astronomy (Astropy Collaboration, 2013). This work has made extensive use of IPython \citep{ipython}, Matplotlib \citep{matplotlib}, astroML \citep{astroML}, scikit-learn \citep{scikit-learn}, and TOPCAT \citep[\url{http://www.star.bris.ac.uk/~mbt/topcat/}]{topcat}. This work would have not been possible without the countless hours put in by members of the open-source community all around the world.   
Finally, CFM gratefully acknowledges an ESA Research Fellowship.
\end{acknowledgements}

\bibliography{bibliografy.bib}
\bibliographystyle{aa}

\begin{appendix}
\section{Color-magnitude and color-color diagrams}
\begin{figure}
\includegraphics[width = \hsize]{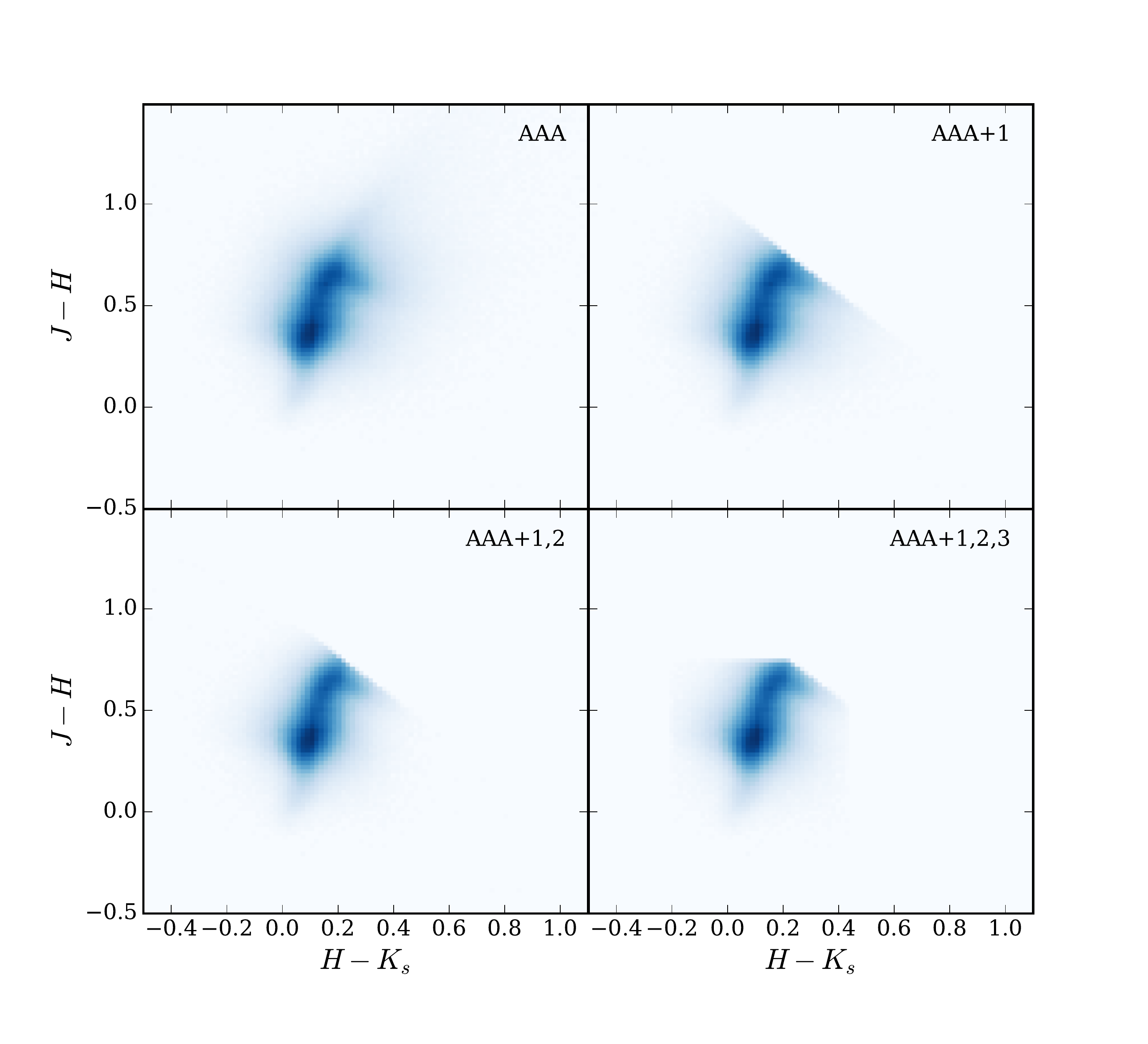}
\caption{Color-color diagrams of the sources resulting from the selection criteria in Section 2.}
\label{fig:app1}
\end{figure}

In this Appendix we show the color-color and color-magnitude diagrams constructed combining \textit{Gaia} DR1 and 2MASS photometry. The sources in the first panel  are those remaining after applying the 2MASS photometry quality selection cut ('ph\_qual = AAA'). The other panels show the cuts of Eq. \eqref{eq:cuts}. Note that we did not apply exactly the same photometric criteria as in \cite{Alves2012} because there is probably a typo in their Eq. (1) that causes 0 sources to be selected. However, Fig. \ref{fig:app1} looks similar to their Fig. 4.

\begin{figure}
\includegraphics[width = \hsize]{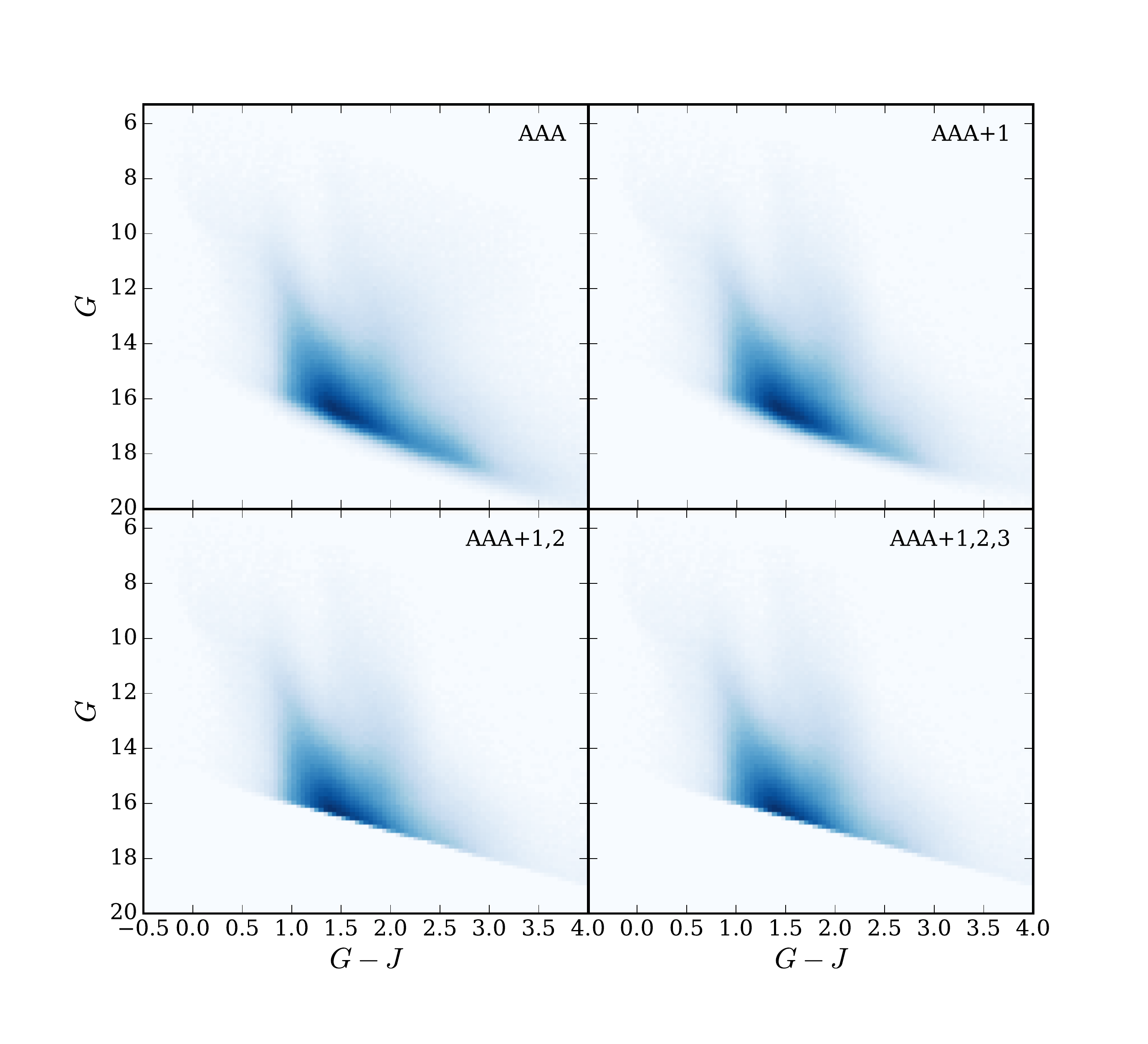}
\caption{Color-magnitude diagrams of the sources resulting from the selection criteria in Section 2.}
\label{fig:app2}
\end{figure}

\onecolumn
\section{ADQL queries}\label{App:ADQL}
We report here the queries used to a) select the sources in our field and b) perform the cross-match with 2MASS. \\
\noindent

\vspace{0.5cm}
\noindent
\textbf{Field selection: \\} 
\noindent
\texttt{select gaia.source\_id, gaia.ra, gaia.dec, gaia.l, gaia.b, gaia.phot\_g\_mean\_mag, gaia.pmra, gaia.pmdec, gaia.parallax, 
gaia.pmra\_error, gaia.pmdec\_error, gaia.parallax\_error  \\
from gaiadr1.gaia\_source as gaia \\
where gaia.l>=190.0 and gaia.l<=220.0 and gaia.b>=-30.0 and gaia.b<=-5.0}

\vspace{0.5cm}
\noindent
\textbf{Cross Match with 2MASS: \\} 
\noindent
\texttt{select gaia.source\_id, gaia.l, gaia.b, gaia.phot\_g\_mean\_mag, 
gaia.phot\_g\_mean\_flux, \\ gaia.phot\_g\_mean\_flux\_error,
gaia.parallax,gaia.parallax\_error, \\gaia.pmra,gaia.pmdec,gaia.pmra\_error,gaia.pmdec\_error, \\
 tmass.j\_m, tmass.j\_msigcom, tmass.h\_m, tmass.h\_msigcom, tmass.ks\_m,  tmass.ks\_msigcom,  \\ tmass.ph\_qual
from gaiadr1.gaia\_source as gaia \\
inner join gaiadr1.tmass\_best\_neighbour as xmatch \\
on gaia.source\_id = xmatch.source\_id \\
inner join gaiadr1.tmass\_original\_valid as tmass \\
on tmass.tmass\_oid = xmatch.tmass\_oid \\
where gaia.l > 190.0 and gaia.l < 220.0 \\and gaia.b < -5.0 and gaia.b > -30.0 and xmatch.angular\_distance < 1.0}

\vspace{0.5cm}
\noindent
We run the queries using the \textit{Gaia} archive. On the archive, we suggest the user to create a personal account. This indeed allows to save queries and store data (up to 1 GB).

\twocolumn
\section{Kernel Density Estimation on the sphere}
The referee pointed out that the kernel density estimation carried out on flat projections of the
Orion sky field will suffer from area distortions, and suggested the use of the von Mises-Fisher
(vMF) kernel, which is intended for analyses on the unit sphere. This kernel is given by the
following equation for a two-dimensional unit sphere (i.e., for three-dimensional unit vectors
$\mathbf{x}$):
\begin{equation}
  f(\mathbf{x}|\mathbf{m}, \kappa) = \frac{\kappa^{1/2}}{(2\pi)^{3/2} I_\frac{1}{2}(\kappa)}
  e^{\kappa\mathbf{m}^\mathsf{T}\mathbf{x}}\,,
  \label{eq:vMFkernel3D}
\end{equation}
where the unit vector $\mathbf{m}$ represents the mean direction for the kernel and
$I_\frac{1}{2}$ is the modified Bessel function of the first kind and order $1/2$, and
$\mathbf{m}^\mathsf{T}\mathbf{x}$ indicates the inner product of $\mathbf{m}$ and $\mathbf{x}$. See
\cite{vmFPaper} for details. The parameter $\kappa\geq0$ indicates the concentration of the
kernel around $\mathbf{m}$. The normalization constant can be re-written by considering that:
\begin{equation}
  I_{\frac{1}{2}}\left(z\right)=\left(\frac{2}{\pi z}\right)^{\frac{1}{2}}\sinh z
\end{equation}
\citep[eq.\ 10.39.1 in][]{NIST}, which leads to:
\begin{equation}
  f(\mathbf{x}|\mathbf{m}, \kappa) = \frac{\kappa}{4\pi\sinh\kappa}
  e^{\kappa\mathbf{m}^\mathsf{T}\mathbf{x}} = \frac{\kappa}{2\pi(e^\kappa-e^{-\kappa})}
  e^{\kappa\mathbf{m}^\mathsf{T}\mathbf{x}}\,.
\end{equation}

The exponent in the kernel contains the inner product of $\mathbf{m}$ and $\mathbf{x}$ and this can
also be written as:
\begin{equation}
  \mathbf{m}^\mathsf{T}\mathbf{x} = \cos\rho = \sin\delta_{m}\sin\delta + \cos(\alpha-\alpha_{m})
  \cos\delta_{m}\cos\delta\,,
\end{equation}
where $(\alpha, \delta)$ represent the ICRS coordinates of the points on the sky, and $\rho$ is the
angle between $\mathbf{m}$ and $\mathbf{x}$. On the unit sphere this angle also represents the
distance along a great circle between the points $\mathbf{m}$ and $\mathbf{x}$, also known as the
`haversine distance'. The value of $\rho$ can also be calculated using the haversine function
($\mathrm{hav}$) given by:
\begin{equation}
  \mathrm{hav}(\theta) = \sin^2\left(\frac{\theta}{2}\right) = \frac{1-\cos\theta}{2}\,.
\end{equation}
The formula for $\rho$ then becomes:
\begin{equation}
  \mathrm{hav}(\rho) = \mathrm{hav}(\delta-\delta_m) + \cos\delta_m\cos\delta \,\mathrm{hav}(\alpha-\alpha_m)\,.
\end{equation}
This can be verified by writing out both sides of the equation in terms of $(1-\cos\theta)/2$.

To continue, we note that the half-width at half maximum of the vMF kernel expressed in terms of
$\rho$ ($\rho_\mathrm{HWHM}$) is given by:
\begin{equation}
  \rho_\mathrm{HWHM} = \arccos\left(1-\frac{\ln 2}{\kappa}\right)\,,
  \label{eq:rhoHwhm}
\end{equation}
where Eq. \ref{eq:rhoHwhm} follows from:
\begin{equation}
e^{\kappa\cos\rho_{\mathrm{HWHM}}} = \frac{e^{\kappa}}{2}, 
\end{equation}
as the maximum of $f(\mathbf{x}|\mathbf{m}, \kappa)$ occurs when $\cos\rho = 1$.\\
\noindent
Equivalently, for a given $\rho_\mathrm{HWHM}$ the corresponding value of $\kappa$ is:
\begin{equation}
  \kappa = \frac{\ln 2}{1-\cos\rho_{HWHM}}\,.
  \label{eq:kappaFromHwhm}
\end{equation}
In our kernel density estimates of source distributions on the sky the kernel sizes are of order 1
degree ($0.017$ radians) or less. This is already in the regime where  to good accuracy
$\cos\rho\approx 1-\rho^2/2$. At the same time the value of $\kappa$ becomes very large ($\sim4550$,
see Eq.\ \ref{eq:kappaFromHwhm} for $\rho_\mathrm{HWHM}=0.017$), such that $\sinh\kappa
\rightarrow \exp(\kappa)/2$. Hence the vMF kernel becomes approximately:
\begin{equation}
  f(\mathbf{x}|\mathbf{m},\kappa) \approx \frac{\kappa}{2\pi}e^{-\frac{\kappa}{2}\rho^2}\,.
\end{equation}
This is in fact a 2D Normal distribution with standard deviations $\sigma=1/\sqrt{\kappa}$ along the
two principal axes, where in the small angle regime one can write $\rho^2 =
(\Delta\alpha\cos\delta)^2 + \Delta\delta^2$, with $\Delta\alpha=\alpha-\alpha_m$ and
$\Delta\delta=\delta-\delta_m$. This shows that in our case (with kernel sizes of a degree or less),
the vMF kernel can be approximated as a 2D Gaussian in terms of the haversine distance.

Our implementation of the kernel density estimate is in Python and makes use of the the
\texttt{sklearn.neighbors.KernelDensity} module in the \texttt{scikit-learn} package by specifying
that the `haversine' metric should be used during the fitting stage of the density estimate (using
the parameters \texttt{kernel='gaussian'} and \texttt{metric='haversine'}).

\end{appendix}
\end{document}